\documentclass[aps,pre,twocolumn,groupedaddress]{revtex4-2}
 
\expandafter\let\csname equation*\endcsname\relax
\expandafter\let\csname endequation*\endcsname\relax
\usepackage{amsmath,amssymb,mathtools}
\usepackage{graphicx,xcolor}
\usepackage{enumerate}

\usepackage[colorlinks,linkcolor=blue,urlcolor=blue,citecolor=blue]{hyperref}
\usepackage[colorlinks,linkcolor=blue,urlcolor=blue,citecolor=blue]{hyperref}
\usepackage{soul,xcolor}
\begin{document}

\title{Ginzburg-Landau amplitude equation for nonlinear nonlocal models}
\author{Stefano Garlaschi}
%\thanks{Equal junior contribution}
\author{Deepak Gupta}
%\thanks{Equal junior contribution}
\author{Amos Maritan}
%\thanks{Equal senior contribution}
\author{Sandro Azaele}
%\thanks{Equal senior contribution}

\address{Dipartimento di Fisica e Astronomia ``Galileo Galilei'', Universit\`a degli Studi di Padova, via Marzolo 8, 35131 Padova, Italy}

\begin{abstract}
{Regular spatial structures emerge in a wide range of different dynamics characterized by local and/or nonlocal coupling terms. In several research fields this has spurred the study of many models, which can explain pattern formation. The modulations of patterns, occurring on long spatial and temporal scales, can not be captured by linear approximation analysis. Here, we show that, starting from a general model with long range couplings displaying patterns, the spatio-temporal evolution of large scale modulations at the onset of instability is ruled by the well-known Ginzburg-Landau equation, independently of the details of the dynamics. Hence, we demonstrate the validity of such equation in the description of the behavior of a wide class of systems. We introduce a novel mathematical framework that is also able to retrieve the analytical expressions of the coefficients appearing in the Ginzburg-Landau equation as functions of the model parameters. Such framework can include higher order nonlocal interactions and has much larger applicability than the model considered here, possibly including pattern formation in models with very different physical features.}

\end{abstract}

%\noindent{\bf Keywords:} {pattern formation, non-linear non-local dynamics, multiple scales analysis.}

\maketitle

%%% To make the table of contents %%%%%
%\noindent\rule{\hsize}{2pt}
%\tableofcontents
%\noindent\rule{\hsize}{2pt}
\markboth{ }{}
%%%%%%%%%%%%%%%%%%%%%%%%%%%%%%%%%%%%%%%

\section{Introduction}

One of the basic mechanisms underpinning the formation of spatial structures is the instability of spatially uniform, and stationary, states under  small perturbations. This simple mechanism is the beginning of pattern formation \cite{cross-book,Hoyle, JD-2, Pismen, Walgraef} and has yielded valuable insights into natural and controlled non-equilibrium systems. The diversity of spatial patterns can be investigated by means of this approach in a wealth of systems, ranging from the archetypal Rayleigh-B\'{e}nard convection \cite{chandrasekhar2013hydrodynamic,cross1993pattern,platten2012convection,di1981instabilities} to reaction-diffusion systems \cite{rd-pat,turing1990chemical,bansagi2011tomography,castets1990experimental,ouyang1991transition}. These latter include reactions of chemical species, eventually leading to regular patterns in coats and skins of animals \cite{murray2002mathematical,nakamasu2009interactions} or seashells \cite{shell}. 

In the evolution equation, an essential role is played by the nonlinear terms that are able to stabilise the initial growth of perturbations and eventually select the spatial pattern. In many examples of interest, including those we have alluded to above, nonlinearities are assumed to be local, albeit spatial patterns can be generated by more general forms of nonlinear terms. 
For instance, the Phase Field Crystal (PFC) theory incorporates crystalline details on length and time scales of experimental relevance and is used to model the structure of several materials \cite{ph-fd,huang}. The connection to the microscopic details is achieved via the Dynamic Density Functional (DDF) theory, from which it can be derived \cite{archer}. In the DDF theory the pairwise and higher order spatial correlation functions are responsible for the nonlocal (and nonlinear) contributions, which govern the evolution of the conserved order parameter.

Several other examples in ecology include the distributions of vegetation as a regular  alternation of colonized regions and bare soil, over the landscapes in many different areas around the globe \cite{rietkerk2008regular, tigerbush, veg-1, veg-2, veg-3}. Interestingly, models describing plant-species dynamics \cite{barbier,prl-veg,fer,couteron, circle-science, Meron-book,Borgogno2009MathematicalMO, shnerb-veg, shnerb-veg-2} provide, to some extent, the physical insights about the origin of such observations. In fact, these models take into account the interactions in the system via nonlocal contributions in the evolution equations, and shed light on the empirical observations interpreting them as pattern formation phenomena. Moreover, they also help in understanding how regular structures over long scales can emerge even in the absence of any environmental perturbation.  

Further, the nonlocal features also play an important role while modelling population dynamics. Herein, the intertwining combination of competition and environmental effects is usually modelled by assuming that species undergo a diffusion process and interact nonlocally in space. Such contributions play a vital role in describing the aggregation and distribution of individuals or species in terms of emerging patterns \cite{pop-1, pop-2,pop-3}. 

Similar settings also enhance our understanding of species origination \cite{hardinscience}. In particular, the competition can indeed lead to formation of species by limiting their similarity and partitioning environmental resources \cite{macarthur1967limiting}. In this case the diffusive process and inter-species interactions occur in the space of species traits, and the eventual patterns obtained from such models are a hallmark of the surviving species \cite{pigolotti2007species,scheffer2006self, leimar}.

The simplest method to have an insight into pattern formation is the linear stability analysis. Within this framework, we gain understanding of the modes which drive instability, and therefore, determine length and time scales that characterize the spatial structures. Typically, these structures are distorted over either large length or large temporal scales, and these slow changes unfortunately cannot be determined by a simple linear analysis. However, near the onset of a supercritical instability \cite{cross1993pattern} and in the weakly nonlinear regime, it is possible to deduce the evolution equation of the amplitude of the most unstable modes, which captures the basic information about those distortions and their relative scales.

Such equation known as the {\it Ginzburg-Landau} (GL) amplitude equation has been obtained first in simple settings like the Rayleigh-B\'{e}nard convection \cite{first-time-GL-1, first-time-GL-2} or the celebrated \textit{Swift-Hohenberg} model \cite{cross1993pattern}. In the following, those results have been extended to several models generating patterns from local dynamics \cite{amp-1,amp-2,amp-3,amp-4, doelman}.
However, the majority of studies taking into account nonlocal features were limited to particular cases, such as the generalizations of the aforementioned Swift-Hohenberg model \cite{kuehn2018validity,morgan2014swift} or the paradigmatic Fisher-KPP equation \cite{nlfish-1, nlfish-2}. In these studies, the authors considered specific settings in order to derive the amplitude equation with nonlocal interaction terms. Thus, to the best of our knowledge, the validity of the amplitude equation in describing the large-scale properties of patterns emerging from a general nonlinear and {nonlocal model has still not been explored.

 In this paper, we focus on this latter problem for systems with nonlinear and nonlocal dynamics exhibiting supercritical instability \cite{cross1993pattern}. Moreover, we assume that the nonlocal couplings are even functions and can be expanded in Taylor series. In this case, we first obtain the criterion for pattern formation in a general model [see Eq.~\eqref{eq: general-eq}]. Then, we obtain the equation that takes the form of the GL equation using a novel mathematical approach based on the expansion of nonlocal operators in the parameter space around the onset of instability. We also show that, near the supercritical onset of instability, where stable pattern solutions emerge continuously from the homogeneous state, the amplitude equation does not depend on the details of the specific model. In other words, we show that the amplitude equation is independent of the form of the nonlinearity and the interaction kernel as long as its Fourier transform exists. Finally, we emphasize that the GL equation depends on the model only through its coefficients [see Eq.~\eqref{eq: amplitude-eq}]. These latter are obtained analytically from the general setting we adopted in our derivation.

The rest of the paper is organized as follows. We first present our general setting in Sec.~\ref{setup}, whereas the mechanism that describes the emergence of patterns is discussed in Sec.~\ref{mech}. Sec.~\ref{nlf-sec} contains the information of the model that we use to illustrate our theoretical formalism. In Sec.~\ref{amp-eqn-sec}, we derive the amplitude equation, and the predicted evolution is compared with numerical simulation in Sec.~\ref{simul}. Finally, we conclude our paper in Sec.~\ref{conc}. Some detailed derivations are relegated in Appendix~\ref{der-28}, Appendix~\ref{der-sec-order}, and Appendix~\ref{third-order}. Some particular solutions of the amplitude equation are shown in Appendix~\ref{P-S}. In Appendix~\ref{num-met}, we discuss the method to obtain the amplitude in numerical simulations. 

%%%%%%%%%%%%%%%%%%%%%%%%%%%%%%%%%%%%%%%%%%%%%%%%%%%%%%%%%%%%%%%%%%%%%%%%%%%%%%%%%%%%%%%%%%%%%%

\section{Problem Setup}
\label{setup}

In this paper, we investigate pattern formation in systems whose evolution is characterized by a nonlocal and nonlinear dynamics in the supercritical regime \cite{cross1993pattern}.
%\sout{a general non-local non-linear dynamics} \textcolor{orange}{a unspecified non-linearity taking into account a symmetric non-local coupling term} {\color{red} again I'd be more specific about our assumptions}.
For the sake of simplicity, we study the dynamics of a real field $\phi(x,t)$, which is governed by the following equation in one spatial dimension
\begin{align}
\partial_t  \phi(x,t)=F_{\textbf{q}}\left[\phi(x,t),\left(G_{\textbf{q}}*\phi\right)(x,t)\right]+D\partial_x^2 \phi(x,t),
\label{eq: general-eq}
\end{align}
 where $F_{\textbf{q}}(\cdot,\cdot)$ is an analytic nonlinear function, $\textbf{q}$ indicates a set of parameters and $D$ a diffusion constant. In the above Eq.~\eqref{eq: general-eq}, for convenience, we write $\partial_y$ for a partial derivative with respect to $y$. Notice that the nonlocal contribution to the equation comes from the convolution of the field with a smooth function $G_{\textbf{q}}(\cdot)$, that plays the role of a kernel, defined as 
\begin{equation}
    (G_{\textbf{q}}*\phi)(x,t)=\int_{-\infty}^{+\infty}~G_{\textbf{q}}( x-y )~\phi(y,t)~dy.
\end{equation}
Moreover, we assume that $G_{\textbf{q}}(\cdot)$ is even, and this function and its Fourier transform can be expanded using the Taylor series. We stress that in our formulation, we are not considering the contribution from the spatial boundaries. Therefore, we can perform the integral over the $x$-variable from $-\infty$ to $+\infty$. he generalization to spatial higher dimensions is straightforward, as long as the kernel maintains the same symmetry properties, e.g., $G(\vec{\textbf{x}})=G(\vert \vec{\textbf{x}} \vert)$. Further, we emphasize that Eq.~\eqref{eq: general-eq} generalizes several models, including the competitive Lotka-Volterra equation \cite{pigolotti2007species,sim-pigg-2, fort-1} and some reaction-diffusion models \cite{turing1990chemical, fishereq, KPP}.
% \textcolor{red}{Any comment on the more general case of a non-symmetric kernel?} 

\section{Mechanism of the emergence of patterns}
\label{mech}
As stated in the Introduction, the patterns start emerging due to the instability of the homogeneous and stationary solution  $\phi^{(0)}_{\textbf{q}}$ and that solution satisfies [See Eq.~\eqref{eq: general-eq}] 
\begin{equation}
    F_\textbf{q}[\phi^{(0)}_{\textbf{q}},\tilde{G}_{\textbf{q}}(k=0)~\phi^{(0)}_{\textbf{q}}]=0,
\end{equation} 
where $\tilde{G}_{\textbf{q}}(k)=\int_{-\infty}^{+\infty}~dz~G_{\textbf{q}}(z)~e^{i k z}$ is the Fourier transform of $G_{\textbf{q}}$, and $k$ being the wavenumber. Spatial patterns that form in the weakly nonlinear regime can be investigated in the region of instability around $\phi^{(0)}_{\textbf{q}}$. Therefore, we substitute 
\begin{align*}
\phi_k(x,t)=\phi^{(0)}_{\textbf{q}} +\delta~e^{\lambda_\textbf{p}(k)t + ikx}+c.c.
\end{align*}
into Eq.~\eqref{eq: general-eq}. Now we assume that the spatially harmonic perturbation is uniformly small; namely, $0<\delta\ll 1$. Thus, up to first order in $\delta$, the growth rate $\lambda_{\textbf{p}}(k)$ as a function of wave number $k$ reads 
\begin{equation}
\lambda_{\textbf{p}}(k)=(1, \tilde{G}_{\textbf{q}}(k)) \cdot \nabla F_{\textbf{q}}\big\vert_{\left(\phi^{(0)}_\textbf{q},\tilde{G}_{\textbf{q}}(0)\phi^{(0)}_\textbf{q}\right)}-D~k^2,
\label{eq: eigenvalue}
\end{equation}
where $\textbf{p}\equiv\{\textbf{q},D\}$ refers to the set of all  parameters of the model and \begin{equation}
 \nabla F_{\textbf{q}}\big\vert_{(x^*,y^*)}=\big[\partial_x F_{\textbf{q}}(x,y)\vert_{(x^*,y^*)}, \partial_y F_{\textbf{q}}(x,y)\vert_{(x^*,y^*)}\big]^\top.   
\end{equation}
Since we assume that $G_{\textbf{q}}(x)$ is an even function, the quantity $\lambda_{\textbf{p}}(k)$ is a real function of $k$.

The stability of $\phi_{\textbf{q}}^{(0)}$ depends on the sign of $\lambda_{\textbf{p}}(k)$, i.e., the homogeneous stationary solution is stable if $\lambda_{\textbf{p}}(k)<0$ for all $k$; otherwise, $\phi_{\textbf{q}}^{(0)}$ is an unstable solution. In fact, the stability of $\phi_{\textbf{q}}^{(0)}$ depends on the system parameters $\textbf{p}$. Therefore, we can find regions in the parameter space to indicate the stability of the solution. Let us call $k_M(\textbf{p})$, a solution of \begin{align*}
\dfrac{\partial\lambda_{\textbf{p}}(k)}{\partial k}\bigg|_{k=k_M(\textbf{p})}=0,
\end{align*} 
a point where the growth rate achieves maximum i.e., $\lambda_{M}(\textbf{p})=\lambda_{\textbf{p}}(k_M(\textbf{p}))$, where the subscript $M$ refers to the maximum.  Notice that both $\lambda_{\textbf{p}}(k)$ and $k_M(\textbf{p})$ are parameterized by system parameters $\textbf{p}$.  Thus, a sufficient condition that the parameters have to fulfill in order to observe pattern formation is $\lambda_M(\textbf{p})>0$. Therefore, in the parameter space a \textit{critical hypersurface} $\mathcal{M}$ can be obtained by setting $\lambda_M\equiv\lambda_{\textbf{p}_0}(k_M(\textbf{p}_0))=0$ where $\textbf{p}_0\equiv\{\textbf{q}_0,D_0\}$ belongs to $\mathcal{M}$, and this hypersurface distinguishes the regions depending on the stability of  $\phi_{\textbf{q}}^{(0)}$.

\section{Example}
\label{nlf-sec}
In order to make our formalism more transparent, we consider the extended Fisher-KPP (F-KPP) equation \cite{fishereq, KPP}, where we also introduce a nonlocal contribution \cite{ref-1-report,ref-2-report,ref-3-report}. We refer to such equation as the nonlocal F-KPP equation.  Notice that this latter is known as nonlocal Lotka-Volterra equation in the ecological literature \cite{pigolotti2007species}. Within this context, the model describes population dynamics characterized by the presence of nonlocal couplings, which can be interpreted as nonlocal interactions of individuals with those that are far away in space or that have different phenotypic traits.

We choose this particular model because it is amenable to analytical calculations and it exhibits pattern forming dynamics in the presence of nonlocal couplings \cite{nlfish-1,nlfish-2}. Therefore, in this example, the first term on the right-hand side of Eq.~\eqref{eq: general-eq} has the following form: 
\begin{equation}\label{NLFF}
F_\textbf{q}[u,v]:=u[1-av],
\end{equation}
where $a$ is a dimensionless parameter. Herein, we consider the functional form of the kernel as following:
\begin{equation}
\label{NLFG}
G_\textbf{q}(z)=\exp\bigg(-\dfrac{|z|}{R}\bigg)-b~\exp\bigg(-\dfrac{|z|}{\beta R}\bigg).
\end{equation}
This form has been chosen mainly because it illuminates the main steps of our calculations for the general model. In Eq.~\eqref{NLFG},  $R$ is the range of the interaction, $\beta$ and $b$ are dimensionless parameters such that $0<b,\beta<1$.

Following Sec.~\ref{setup}, we obtain the the homogeneous and stationary solution as
\begin{align}
    \phi^{(0)}_\textbf{q}=[a\tilde{G}_\textbf{q}(0)]^{-1},
    \end{align}
    where 
    \begin{align}
        \tilde{G}_\textbf{q}(k)=2R\bigg(\dfrac{1}{1+k^2R^2}-\dfrac{b\beta}{1+k^2R^2\beta^2}\bigg).
    \end{align}
Similarly, the dispersion relation using Eq.~\eqref{eq: eigenvalue} can be obtained as
\begin{equation}\label{dispersion}
\lambda_\textbf{p}(k)=\frac{1}{1-b \beta }\bigg(\frac{b \beta }{1+\beta ^2 k^2 R^2}-\frac{1}{1+k^2 R^2}\bigg)-D~k^2,
\end{equation} 
in which $\textbf{p}=\{b,\beta,a,R,D\}$ is the set of parameters as discussed in Sec.~\ref{setup} and $\lambda_\textbf{p}(k)$ does not depend on $a$. We plot $\lambda_\textbf{p}(k)$  {\it vs.} $k$ in the left panel of Fig.~\ref{fig:phase-diag-and-lambda} for three different values of $b$, while the other parameters are kept fixed.

%\onecolumngrid
%\begin{center}
\begin{figure*}
\centering
\includegraphics[width=8 cm]{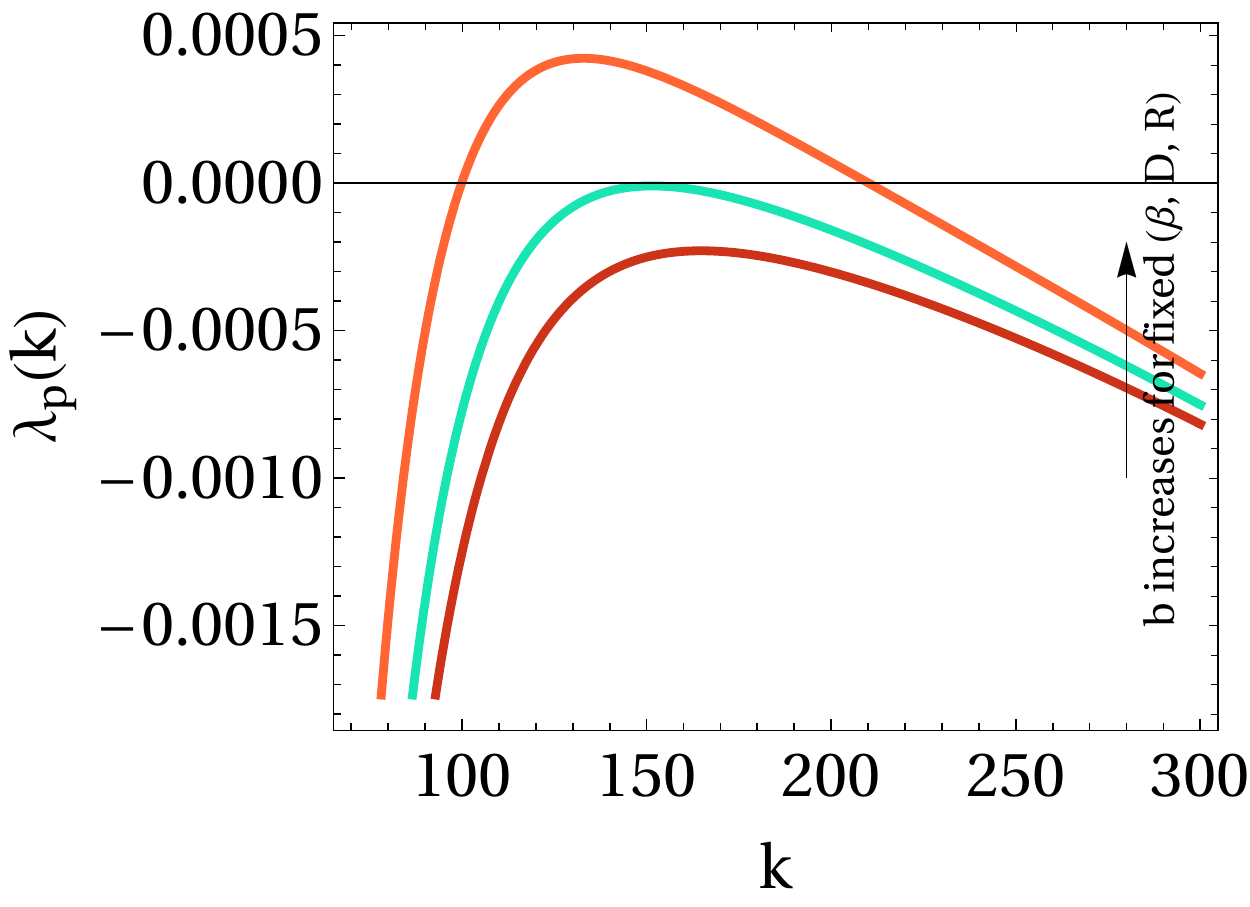}~~~~~~~~~~~~
\includegraphics[width=6 cm]{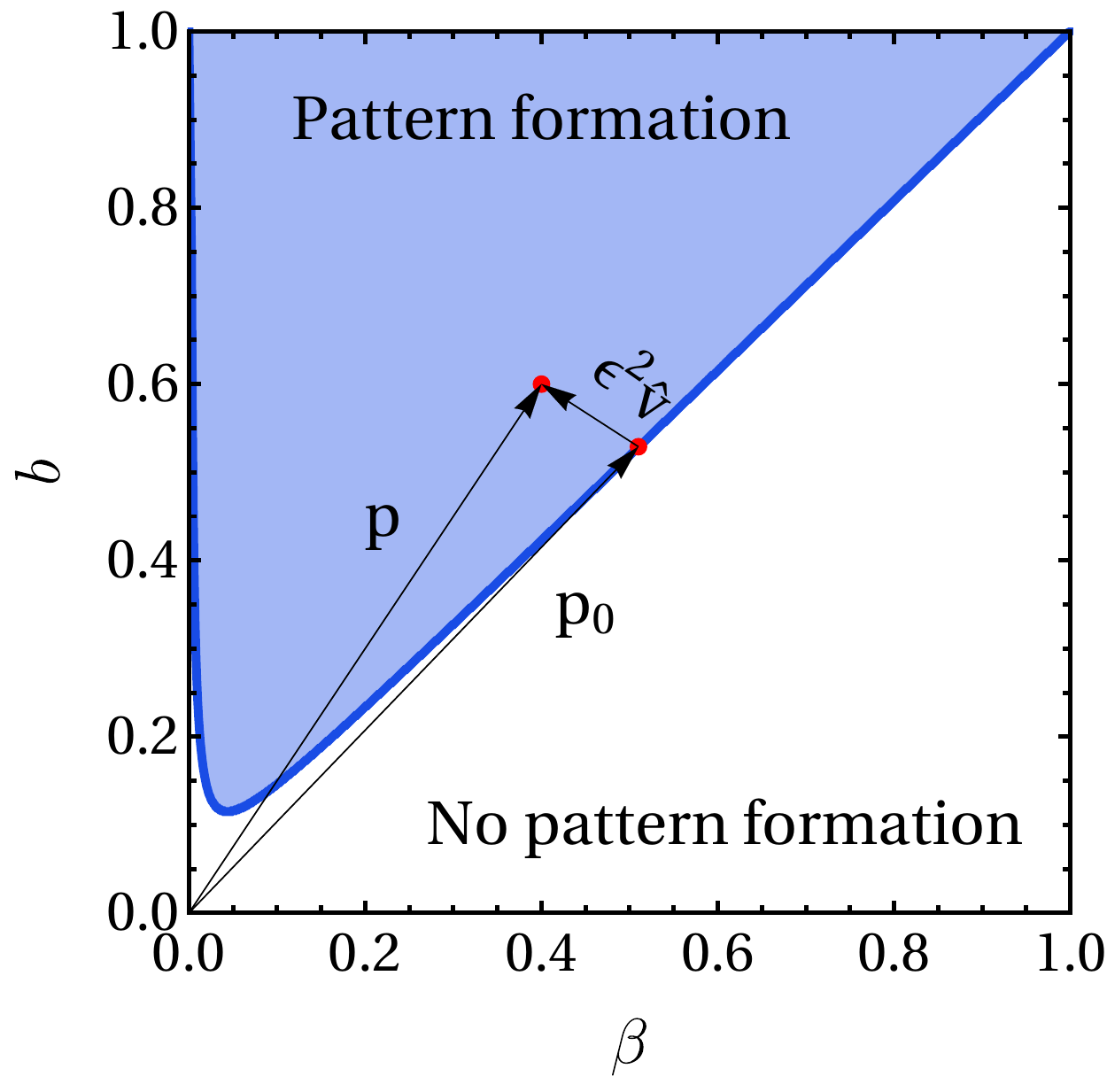}
\caption{Left panel: The dispersion relation given in Eq.~\eqref{dispersion} $\lambda_{\textbf{p}}(k)$ as a function of $k$ for the nonlocal F-KPP equation at three different values of $b$. The remaining parameters for the plots are $\beta=0.2$, $D=10^{-8}$, and $R=0.1$. Right panel: Phase diagram in the ($\beta,b$) space for the nonlocal F-KPP equation given in Eqs.~\eqref{eq: general-eq} and \eqref{NLFG} with $F_\textbf{q}[u,v]:=u[1-av]$. In this case $\textbf{p}\equiv\{\beta, b, R, a,D\}$ and the critical hyper-surface $\mathcal M$ does not depend on $a$. The phase diagram is shown for two fixed parameters $D=10^{-8}$ and $R=0.1$, where the solid contour $\mathcal{M}$ [defined by $\lambda_{\textbf{p}_0}(k_M(\textbf{p}_0))=0$]  divides the parameter space depending on whether or not there is pattern formation. A vector $\textbf{p}=\textbf{p}_0 + \epsilon^2 \hat{v}$ indicates a point in the pattern forming region, where $\textbf{p}_0$ sits on $\mathcal{M}$.}
\label{fig:phase-diag-and-lambda}
\end{figure*} 
%\end{center}
%\twocolumngrid

In order to obtain the phase diagram that identifies the region of stability, we study the sign of maximum of $\lambda_\textbf{p}(k)$ by varying the parameters $\textbf{p}$. Specifically, the critical hypersurface, that divides the parameters space, we obtain by setting such maximum equal to zero. The analytical computation to find this phase boundary  is difficult. Nevertheless, we numerically obtain the phase diagram in the $(\beta,b)$ plane for other fixed parameters, and it is shown in Fig.~\ref{fig:phase-diag-and-lambda}(right panel), where the blue shaded region indicates the region of instability of the homogeneous and stationary solution. Thus, we name that region as pattern forming region.

\section{Amplitude equation}
\label{amp-eqn-sec}
This section is dedicated to the derivation of the amplitude of the pattern near the contour of instability in the general case of which Fig.~\ref{fig:phase-diag-and-lambda}(right panel) is a particular case.

In order to make analytical progress, we use the Taylor series expansion  of the  right-hand  side of Eq.~\eqref{eq: general-eq} around the homogeneous and stationary solution $\phi^{(0)}_{\textbf{q}}$,  i.e., we expand the nonlinear function $F_{\textbf{q}}(\cdot,\cdot)$ around $\left(\phi^{(0)}_{\textbf{q}},\tilde{G}_{\textbf{q}}(0)~\phi^{(0)}_{\textbf{q}}\right)$. This allows to set up
equations that hold in the weakly nonlinear regime and finally obtain the amplitude equation. We express the field as $\phi(x,t)=\phi^{(0)}_{\textbf{q}}+\varphi(x,t)$. 
%, and consider $\varphi(x,t)$ to be small compared to $\phi^{(0)}_{\textbf{q}}$, i.e., $|\varphi(x,t)|\ll |{\color{red}\phi^{(0)}_{\textbf{q}}}|$.  
The evolution equation for $\varphi(x,t)$ can then be cast in the form:
\begin{align}
\dot{\varphi}=\mathcal{L}_{\textbf{p}}\varphi+\mathcal{N}_{\textbf{q}}\varphi,
\label{eq: expand-eq}
\end{align}
where the first and second term, respectively, on the right-hand side correspond to linear and nonlinear contributions in $\varphi$. In the above equation \eqref{eq: expand-eq}, the linear operator has the following structure
\begin{align}
\mathcal{L}_\textbf{p}\varphi&=\left(\varphi, G_\textbf{q} * \varphi\right) \cdot \nabla F_\textbf{q}\vert_{\left(\phi_{\textbf{q}}^{(0)}\tilde{G}_{\textbf{q}}(0)\phi_{\textbf{q}}^{(0)}\right)}+D \partial_x^2 \varphi\notag\\&=C^{(1,0)}_{\textbf{q}}\varphi+C^{(0,1)}_{\textbf{q}}\left(G_\textbf{q} * \varphi\right)+D\partial_x^2 \varphi,
\label{eq: exp-local-op}
\end{align}
while the nonlinear operator is
\begin{equation}
\mathcal{N}_{\textbf{q}}\varphi=\sum_{\substack{n,m=0\\ \text{with } n+m \geq 2}}^{+\infty} C^{(n,m)}_{\textbf{q}} \varphi^n  \left(G_{\textbf{q}} * \varphi\right)^m,
\label{eq: exp-non-local-op}
\end{equation}
where $C^{(n,m)}_{\textbf{q}}$ are the coefficients obtained from the Taylor series expansion.

We notice that Eq.~\eqref{eq: general-eq} is translational invariant. Therefore, the eigenfunctions of the linear nonlocal operator $\mathcal{L}_{\textbf{p}}$ are the simple wavefunctions $e^{ikx}$, and then, the eigenvalue equation reads 
\begin{align}
    \mathcal{L}_{\textbf{p}}e^{ikx}=\lambda_{\textbf{p}}(k)~e^{ikx},
\end{align} 
where the spectrum is defined in Eq.~\eqref{eq: eigenvalue}. The general solution of the linear part of Eq.~\eqref{eq: expand-eq}, i.e., $\partial_t{\varphi}(x,t)=\mathcal{L}_{\textbf{p}}\varphi$, is a linear combinations of functions $e^{\lambda_\textbf{p}(k)t + ikx}$ with $k$ dependent coefficients. In this case, Eq. \eqref{eq: eigenvalue} becomes 
\begin{equation}
\lambda_\textbf{p}(k)=C^{(1,0)}_{\textbf{q}}+C^{(0,1)} \tilde{G}_{\textbf{q}}(k)-Dk^2.\label{lamb-2}
\end{equation}
To illuminate Eq.~\eqref{eq: expand-eq}, we again consider our model discussed in Sec.~\ref{nlf-sec}. Herein, the linear operator acting on the perturbation field $\varphi$ has the following form:
\begin{align}
    \mathcal{L}_{\textbf{p}}\varphi=- [\tilde{G}_\textbf{q}(0)]^{-1}(G_{\textbf{q}}* \varphi)+D\partial_x^2\varphi,
    \end{align}
    and the second term on the right-hand side of Eq.~\eqref{eq: expand-eq} can be shown as 
    \begin{align}
    \mathcal{N}_{\textbf{q}}\varphi=-a~\varphi(G_{\textbf{q}} * \varphi).
    \end{align}
    
In what follows, unless specified, we focus on our general setting described in Eq.~\eqref{eq: general-eq}. 

To obtain the equation that describes the evolution (whose form will be discussed later) of the patterns near the bifurcation contour, we investigate the behavior of the system close to the onset of instability, namely near the critical hyper-surface $\mathcal{M}$. Thus, we consider parameters $\textbf{p}$ in the neighborhood of  $\textbf{p}_0\equiv\lbrace \textbf{q}_0, D_0 \rbrace$, i.e., 
\begin{equation}
\textbf{p}=\textbf{p}_0 + \epsilon^2 \hat{v},
\label{eq: par-exp}
\end{equation}
where $\textbf{p}_0 \in \mathcal{M}$, $\hat{v}$ is a unit vector  pointing toward the region of pattern formation, and $0<\epsilon^2\ll 1$. An example of such point \textbf{p} for nonlocal F-KPP equation (see Sec.~\ref{nlf-sec}) is indicated in the left panel of Fig.~\ref{fig:phase-diag-and-lambda}. 

In addition, we assume that the growth rate $\lambda_{\textbf{p}}(k)$ exhibits a quadratic scaling in the wave-number $k$ close to the point of maximum $k_M(\textbf{p})>0$, which is satisfied if $\lambda_{\textbf{p}}(k)$ admits continuous second derivative with respect to $k$.

With a set of parameters $\textbf{p}$ that can be expressed as in Eq.~\eqref{eq: par-exp} with $\epsilon$ small, we can expand the growth rate around $\textbf{p}_0$ as
\begin{equation}
\lambda_{\textbf{p}}(k)=\lambda_{\textbf{p}_0}(k)+\epsilon^2\hat{v} \cdot \nabla_{\textbf{p}} \lambda_{\textbf{p}}(k)\vert_{\textbf{p} = \textbf{p}_0}+\mathcal{O}\left(\epsilon^4\right),
\label{eq: exp-lambda-p}
\end{equation}
where we assume that the second term on the right-hand side is non-zero.

We know that the above function achieves the maximum at $k=k_M(\textbf{p})$, and that $k_M(\textbf{p})$ can also be expanded about $\textbf{p}_0$
\begin{equation}
k_M(\textbf{p})=k_M(\textbf{p}_0)+\epsilon^2\hat{ v} \cdot \nabla_{\textbf{p}} k_M\vert_{\textbf{p}=\textbf{p}_0}+\mathcal{O}\left(\epsilon^4\right).
\label{eq: exp-kM}
\end{equation}
Substituting Eq. \eqref{eq: exp-kM} in Eq. \eqref{eq: exp-lambda-p} at $k=k_M(\textbf{p})$, we get 
\begin{align}
\lambda_M&\equiv\lambda_{\textbf{p}}(k_M(\textbf{p}))\notag \\
&=\lambda_{\textbf{p}_0}(k_M(\textbf{p}))+\epsilon^2\hat{v} \cdot \nabla_{\textbf{p}} \lambda_{\textbf{p}}(k_M(\textbf{p}))\vert_{\textbf{p}=\textbf{p}_0}~+~\mathcal{O}\left(\epsilon^4\right)\notag\\ 
&=\underbrace{\lambda_{\textbf{p}_0}(k_M(\textbf{p}_0))}_{=0}~+~\epsilon^2\hat{v}\cdot \nabla_{\textbf{p}} k_M\vert_{\textbf{p}=\textbf{p}_0}\underbrace{\lambda'_{\textbf{p}_0}(k_M(\textbf{p}_0))}_{=0}+ \notag \\
&+\epsilon^2\underbrace{\hat{ v}\cdot \nabla_{\textbf{p}} \lambda_{\textbf{p}}(k_M(\textbf{p}_0))\vert_{\textbf{p}=\textbf{p}_0}}_{\bar{\lambda}_M}
+\mathcal{O}\left(\epsilon^4\right).
\label{Seq: exp-lambda-max-tot}
\end{align}
Therefore, we find that the maximum scales like $\epsilon^2$ as $\epsilon\rightarrow 0^+$, i.e.,
\begin{equation}
\lambda_M \to \epsilon^2 \bar{\lambda}_M \quad \text{as }\qquad\epsilon \to 0^+,
\label{Seq: scaling-lambda}
\end{equation}
where we introduce the re-scaled quantity $\bar{\lambda}_M$, which is $\mathcal{O}(1)$.

Owing to this scaling property, we can introduce a temporal- and spatial-scale separation which simplifies Eq.~\eqref{eq: expand-eq}. The long time modulations of the fast oscillations evolve on scales determined by the slower time variable $\tau= \epsilon^2 t$. A similar spatial-scale separation for the perturbation field $\varphi(x,t, \epsilon)$ occurs with a spatial scale given by the slower variable $\xi=\epsilon x$.  Therefore we make the educated guess that the $\epsilon$ dependence is as follows: $\varphi(x, \xi, t)=\sum_{j\geq1} \epsilon^j \varphi_j (x,\xi,\tau)$ where the time dependence in each mode on the right-hand side is through $\tau$. Similarly the spatial dependence appears both through the $x$ and the slower variable $\xi$ \cite{Hoyle}.

Due to these separation of scales, the time derivative transforms as
\begin{equation}
\partial_t \to \epsilon^2 \partial_\tau,\label{eq: time-sc}
\end{equation}
while the spatial derivative encoded in the linear operator becomes
\begin{equation}
\partial_x \to \partial_x+\epsilon \partial_\xi.
\label{eq: space-der}
\end{equation}
As discussed above, $\varphi(x,\xi, \tau)$ can be written as a power series in $\epsilon$, i.e.,
\begin{equation}
\varphi(x,\xi, \tau)=\sum_{i\geq 1} \epsilon^i \varphi_i(x,\xi, \tau),
\label{eq: perturb}
\end{equation}
 From the above expression \eqref{eq: perturb}, we see that close to the bifurcation, only first terms will be dominant and that will determine the growth of the patterns. 

Similar to Eqs. \eqref{eq: exp-lambda-p} and \eqref{eq: exp-kM}, we also expand the linear and nonlinear operators appearing in Eqs.~\eqref{eq: exp-local-op} and \eqref{eq: exp-non-local-op}:
\begin{equation}\label{lin-op}
\mathcal{L}_{\textbf{p}} =\mathcal{L}_{\textbf{p}_0}+\epsilon^2 \overbrace{\hat{v}\cdot \left(\nabla_{\textbf{p}}\mathcal{L}_{\textbf{p}}\right)\vert_{\textbf{p}=\textbf{p}_0}}^{\delta \mathcal{L}_{\textbf{p}_0}}+\mathcal{O}\left(\epsilon^4\right),
\end{equation}
\begin{equation}
\mathcal{N}_{\textbf{q}} =\mathcal{N}_{\textbf{q}_0}+\epsilon^2 \hat{v}\cdot \left(\nabla_{\textbf{p}}\mathcal{N}_{\textbf{q}}\right)\vert_{\textbf{p}=\textbf{p}_0}+\mathcal{O}\left(\epsilon^4\right) \label{eq: nl-exp}. 
\end{equation}

Next, we proceed as follows. We first substitute Eqs. \eqref{eq: time-sc}--\eqref{eq: nl-exp} into Eq. \eqref{eq: expand-eq}, and then we introduce  the spatial scale separation in $\mathcal{L}_{\textbf{p}_0}$ and in the nonlocal terms of $\mathcal{N}_{\textbf{q}_0}$ (See Appendix~\ref{der-28} for detailed derivation). Finally, we arrive at
\begin{align}
\epsilon^3 \dot{\varphi_1} + o\left(\epsilon^3\right)&=\epsilon H_1(\textbf{p}_0,\varphi_1)+\epsilon^2H_2(\textbf{p}_0,\varphi_1,\varphi_2)+\notag \\
&+\epsilon^3 H_3(\textbf{p}_0,\varphi_1,\varphi_2),
\label{eq-all-orders-with-space}
\end{align}
where the functional form of $H_i$ is given in Appendix~\ref{der-28}, and we remind that $\textbf{p}_0\equiv\{\textbf{q}_0,D_0\}$.

The above equation \eqref{eq-all-orders-with-space} is the starting point to obtain the amplitude equation. To proceed further, as a standard approach, we will compare the coefficients on the left and right-hand side of the equation at same order in $\epsilon$. Let us first begin with the first order contribution. At the lowest order in $\epsilon$, we find from Eq.~\eqref{eq-all-orders-with-space} that

\begin{equation}
H_1(\textbf{p}_0, \varphi_1)=0.
\end{equation}
Thus, from the expression of $H_1(\textbf{p}_0, \varphi_1)$ shown in Appendix~\ref{der-28} one can easily write the solution of this equation as:
\begin{equation}
\varphi_1(x,\xi, \tau)=A(\xi,\tau)~e^{ik_M(\textbf{p}_0) x} + \bar{A}(\xi, \tau)~e^{-i k_M(\textbf{p}_0) x}\label{eq-phi-1}.
\end{equation}
The functional form of $\varphi_1(x,\xi, \tau)$ suggests that it has harmonic oscillation with the mode characterized by $k_M(\textbf{p}_0)$. We further notice that, the temporal dependence is only present through the amplitude of this harmonic oscillation on a slower scale defined by $\tau$. Moreover, such amplitude may display a spatial evolution, but on the longer scale given by $\xi$. Near criticality, we expect that this is the relevant contribution to the pattern formation. Thus, to understand the growth of the patterns near bifurcation, we aim to obtain the equation for that amplitude.

Next, we compare the second order contribution $\mathcal{O}(\epsilon^2)$ in Eq.~\eqref{eq-all-orders-with-space}, and then, use the first order solution \eqref{eq-phi-1}, we obtain (see Appendix~\ref{der-sec-order} for details) 
\begin{align}
\varphi_2(x,\xi,\tau)&=\overbrace{B(\xi,\tau)e^{ik_M(\textbf{p}_0) x} + \bar{B}(\xi,\tau)e^{-i k_M(\textbf{p}_0) x}}^{\Lambda(x,\xi,\tau)}+ \notag \\
&+\Sigma_{\textbf{p}_0}\bigg[\frac{A^2(\xi,\tau) e^{2ik_M(\textbf{p}_0) x}}{\lambda_{\textbf{p}_0}(2k_M(\textbf{p}_0))}+2\frac{\vert A \vert ^2(\xi,\tau)}{\lambda_{\textbf{p}_0}(0)}+\notag \\
&+\frac{\bar{A}^2(\xi,\tau) e^{-2ik_M(\textbf{p}_0) x}}{\lambda_{\textbf{p}_0}(2k_M(\textbf{p}_0))}\bigg],
\label{sol-second order}
\end{align}

Note that the system is at the onset of bifurcation, and we have $\vert \epsilon^2 \varphi_2(x,\xi,\tau) \vert \ll \vert \epsilon \varphi_1(x,\xi,\tau) \vert$. Therefore, due to the choice of the parameters, $\varphi_2(x,\xi,\tau)$ does not play any significant role in shaping the patterns. Hence, Eq.~\eqref{eq-phi-1} would be sufficient to predict the patterns characterized by the amplitude $A(\xi,\tau)$.

%\onecolumngrid
%\begin{center}
\begin{figure*}
\centering 
\includegraphics[width=8cm]{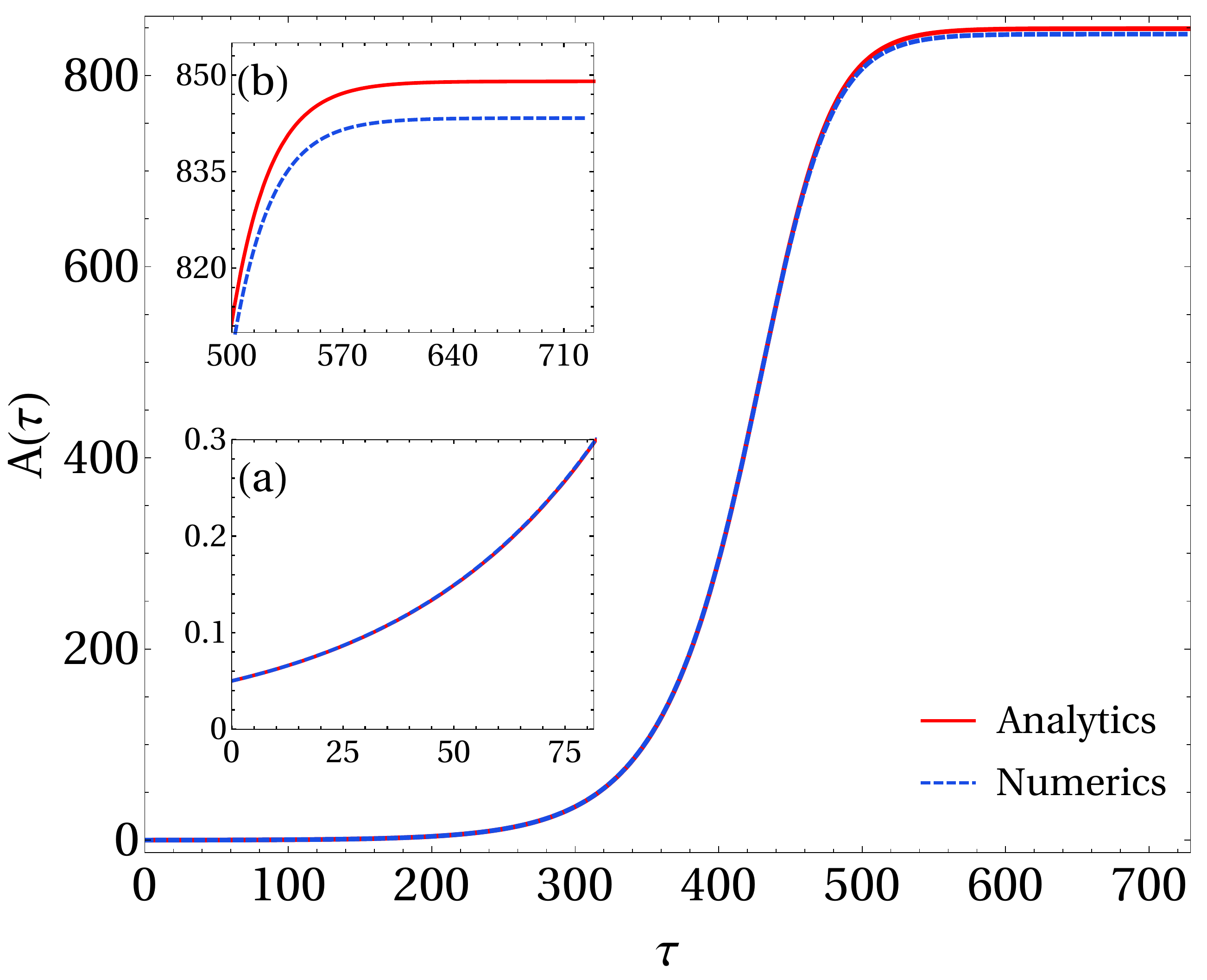}\qquad
\includegraphics[width=8cm]{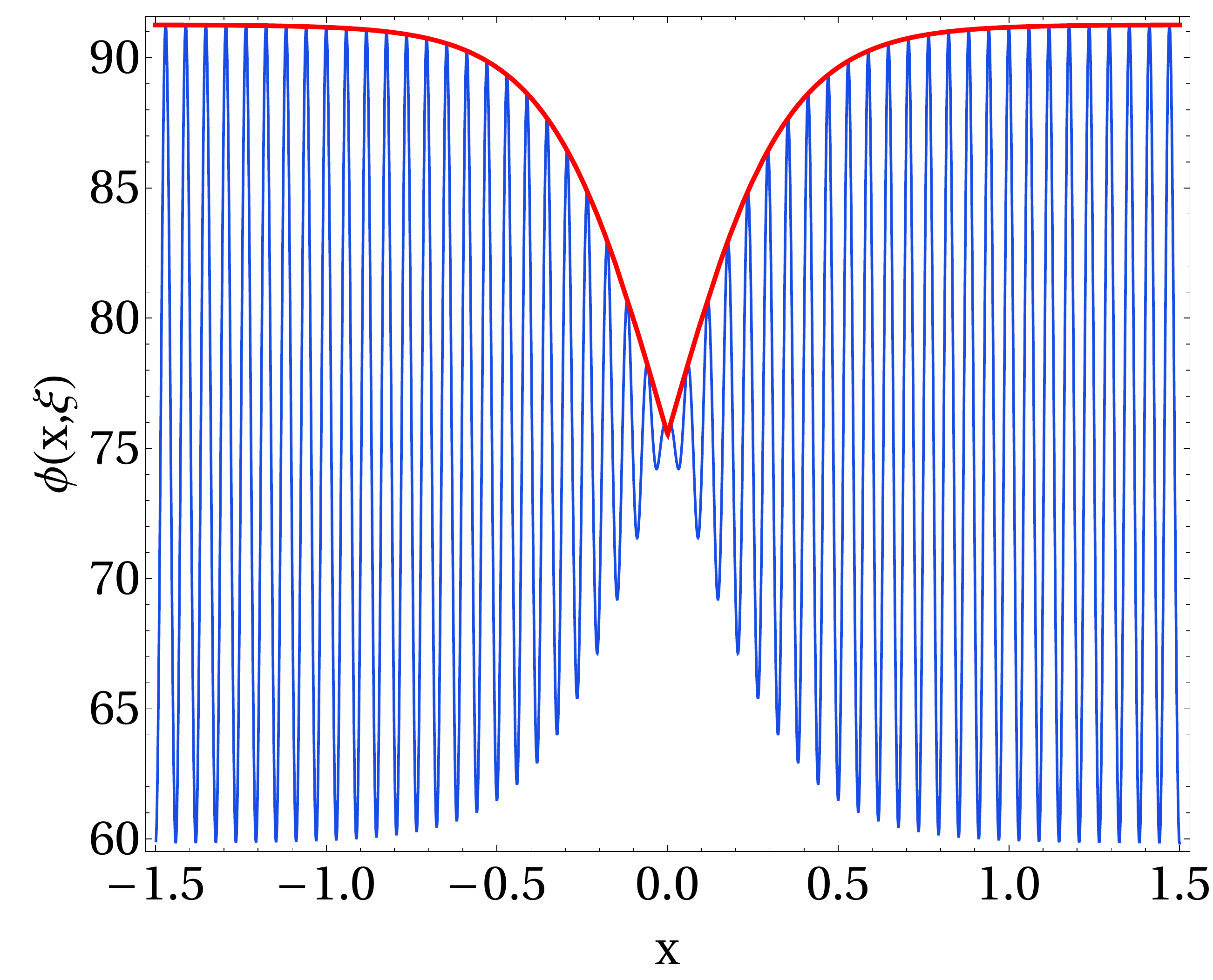}
\caption{Left panel: Comparison between the growth in time of the amplitude predicted by Eq.~\eqref{eq: amplitude-eq} from the initial condition $A(\xi, \tau=0)= A_0=0.05$ (solid red line) and the corresponding numerical evaluation (blue dashed line) from the integration of the nonlocal F-KPP equation using $\phi(x,\xi,0)=\phi_{\textbf{q}}^{(0)}+2\epsilon A_0 \cos(k_M(\textbf{p}_0) x)$ as an initial condition (see Appendix~\ref{num-met}). Owing to this choice, the amplitude remains space-independent at any time, displaying only temporal changes (see Appendix~\ref{P-S}). We refer to Appendix~\ref{num-met} for the details of the parameters $\textbf{p}$ and $\textbf{p}_0$ used in both analytics and numerical simulation. The insets show the zoom on the initial growth (a) and the saturation observed at large time (b). We can notice a remarkable agreement between two curves at all times. Right panel: Comparison between the spatially-dependent stationary solution of Eq.~\eqref{eq: amplitude-eq}, $A_{\rm{st}}(\xi)$, presented in Appendix~\ref{P-S} (the red solid line is the envelope curve $\phi_{\textbf{q}}^{(0)}+2\epsilon A_{\rm{st}}(\xi)$, where $\xi=\epsilon x$) and the solution obtained from the numerical integration of the nonlocal F-KPP equation using $\phi(x,\xi,0)=\phi_{\textbf{q}}^{(0)}+2\epsilon A_{\rm{st}}(\xi) \cos(k_M(\textbf{p}_0) x)$ as initial condition (see Appendix~\ref{num-met}). This plot is obtained at time $t=10^2$ (time steps). The parameters $\textbf{p}$ and $\textbf{p}_0$ along with a discussion of this solution are included in Appendix~\ref{num-met}. We can appreciate how the carrier wave obtained from the numerical integration shows a remarkable agreement with the analytical solution calculated in the weakly nonlinear regime. This suggests that our framework is able to describe also the spatial modulations of the envelope of the emerging patterns.}
\label{fig: amplitude-2-body}
\end{figure*}
%\end{center}
%\twocolumngrid

Finally, on comparing third order contributions (see Appendix~\ref{third-order} for details) and utilizing the solutions given in Eqs.~\eqref{eq-phi-1} and \eqref{sol-second order}, we obtain the growth equation for $A(\xi, \tau)$:
\begin{equation}
\frac{\partial A}{\partial \tau}=\bar{\lambda}_M A-\alpha \vert A \vert ^2  A+\frac{1}{2}\vert \lambda ''_{\textbf{p}_0}(k_M(\textbf{p}_0)) \vert~\frac{\partial^2 A}{\partial \xi^2},
\label{eq: amplitude-eq}
\end{equation}
where we have dropped the dependence $\xi$ and $\tau$ from $A(\xi,\tau)$. We stress that the above equation \eqref{eq: amplitude-eq} is obtained by ensuring that the higher-order terms in the expansion of Eq.~\eqref{eq: expand-eq} are well defined. In the above equation~\eqref{eq: amplitude-eq}, all coefficients on the right-hand side depend on $\textbf{p}_0$, and the detailed expression of the constant $\alpha$ in terms of model details is given in Appendix~\ref{third-order}. 
 Eq.~\eqref{eq: amplitude-eq} represents our main result, and interestingly, it is the celebrated GL equation for a complex field $A(\xi,\tau)$.

%{\color{red} Since the interaction kernel $G_{\textbf{q}}(\cdot)$ is even (which ensures the dispersion relation Eq.~\eqref{eq: eigenvalue} to be real valued), the resulting amplitude equation \eqref{eq: amplitude-eq} has real coefficients. Relaxing such constraint in the \textcolor{red}{nonlocal} coupling term, one may end up with a complex amplitude equation that can generate more complicated behaviors, including spatio-temporal intermittency and phase turbulence (for example, see Ref.~\cite{complexGL}). More generally, complex amplitude equation might arise when the perturbation growth rate has non zero imaginary part, leading to an oscillatory dynamics at the supercritical onset.} In our analysis, we have considered systems whose interaction kernel is smooth in the weakly \textcolor{red}{nonlinear} regime. Should the coupling be strong, those expansions were not valid \cite{strong-1,strong-2} and a different approach is necessary. We leave this study for a future investigation. 
Since the interaction kernel $G_{\textbf{q}}(\cdot)$ is even, the resulting amplitude equation \eqref{eq: amplitude-eq} has real coefficients. Relaxing such constraint in the nonlocal coupling term, one may end up with a complex amplitude equation that can generate more complicated behaviors, including spatio-temporal intermittency and phase turbulence (for example, see Ref.~\cite{complexGL}). In our analysis, we have considered systems whose interaction kernel is smooth in the weakly nonlinear regime. Should the coupling be strong, those expansions were not valid \cite{strong-1,strong-2} and a different approach is necessary. We leave this study for a future investigation.

In our framework that includes the expansion of nonlocal operators in the parameters space at the onset of instability, we explicitly demonstrate that the GL equation emerges from a larger class of models, irrespective of whether systems have nonlocal interactions or not. In particular, we show that this equation is universal, namely only the three coefficients
of Eq.~\eqref{eq: amplitude-eq} are affected by the specific form of the model defined by Eq.~\eqref{eq: general-eq} (see Appendix~\ref{third-order}). 

For example, when Eq.~\eqref{eq: general-eq} defines a nonlocal F-KPP equation, we retrieve the amplitude equation obtained in \cite{nlfish-1}, in which, however, a slow spatial variable was not included. Instead, if we use the explicit forms of $F$ and $G$ [see Eq.~\eqref{eq: general-eq}] given in Ref.~\cite{nlfish-2}, we exactly end up with Eq.~\eqref{eq: amplitude-eq}.

\section{Numerical simulation}
\label{simul}
 We confirm eq. (32) with the numerical integration of the model discussed in Sec.~\ref{nlf-sec}, i.e., the nonlocal F-KPP equation, obtained inserting Eqs.~\eqref{NLFG} and~\eqref{NLFF} into Eq.~\eqref{eq: general-eq}. For fixed parameters $\textbf{p}$ and $\textbf{p}_0$, we consider two cases, which differ by the choice of the initial conditions used in the amplitude equation as well as for the evolution of the nonlocal F-KPP equation. In the first one, we take a homogeneous initial condition for the amplitude, while in the second we set the initial condition to be a particular stationary solution of Eq.~\eqref{eq: amplitude-eq} (discussed in Appendix~\ref{P-S}). The comparison between analytical predictions and numerical results are shown in Figs.~\ref{fig: amplitude-2-body}(left panel) and~\ref{fig: amplitude-2-body}(right panel). In both figures a remarkable agreement can be observed, suggesting the validity of our findings for temporally and spatially  modulated patterns. The numerical amplitude and the predicted envelope displayed in Fig.~\ref{fig: amplitude-2-body} are obtained by taking into account only the first order term \eqref{eq-phi-1} of the perturbative expansion. In Appendix~\ref{num-met} we present the results for the numerical evaluation of the amplitude when considering the next-to-leading order terms and compare with the numerical simulation, and they also have a very good agreement.

\section{Conclusions}
\label{conc}
In this paper, we have considered a general model which can describe pattern formation in several physical systems. We have combined nonlocal coupling terms and nonlinear interactions, which may possibly include many-body terms. From this dynamics, the patterns can emerge when the homogeneous stationary solution becomes unstable. As an example, we can think of an ecological model defined on the abstract niche space, where species emerge as a trade-off between nonlocal interactions and their tendency to scour the space for better evolutionary solutions.  In this case, we find regularly spaced lumps, showing a general tendency of species to coexist when they are either sufficiently similar or sufficiently different, with typical distance of lumps $\mathcal{O}(k_M^{-1}(\textbf{p}_0))$ along a niche axis.

The amplitude of the patterns emerging from dynamics described by Eq.~\eqref{eq: general-eq} is dictated by the universality which operates near the instability. %{\color{cyan}Deepak: I think we can remove this line. Regarding the instability, we have already written a line above. }
The aforementioned universality is particularly interesting for the implications. The key steps in our derivations -- e.g., the introduction of the nonlocal linear operator $\mathcal{L}_{\textbf{p}}$, the expansion close to the boundaries of the critical hyper-surface $\mathcal{M}$ where a quadratic scaling occurs -- could equally well be applied to models with different physical features. For instance, nonlocal higher-order interactions may play an important role in shaping patterns of many physical systems, e.g., ecological communities, and may also help to stabilize their dynamics \cite{grilli2017higher}. The inclusion of such contributions in our framework is straightforward. One just need to insert in the function $F_\textbf {q}$ in Eq.~\eqref{eq: general-eq} terms with the form
\begin{equation}\label{eq: higher-moved}
\int ~G_{\textbf{q}}( x-y_1,x-y_2,\dots,x-y_n )~\prod_{i=1}^n\big[\phi(y_i,t)~dy_i\big].
\end{equation}
Close to instability, those terms will affect only the coefficients of the GL equation \eqref{eq: amplitude-eq}.  Further, by replacing $F_{\textbf{q}}$ with $\partial_x^2(\delta\mathcal{F}_{\textbf{q}}/\delta \phi)$ in Eq.~\eqref{eq: general-eq}, we could also describe the dynamics of a conserved order parameter as we have alluded to in the Introduction. Large scale modulation of patterns of such fields may still be described by GL equations.
%{\color{red} Finally, generalized GL equations for many amplitudes could be derived for systems with many interacting fields $\phi_m(x,t)$, where $m$ is a discrete index. We expect the number of components in the amplitude equation to be determined by the symmetries and the conservation laws occurring in the system also in the presence of long range coupling terms. This is left for future studies.}
Finally, generalized GL equations for many amplitudes could be derived for systems with many interacting fields/species $\phi_m(x,t)$, with $m$ being a discrete index. We expect that, even in the presence of long range coupling terms, the number of components in the amplitude equation is determined by the symmetries and the conservation laws of the system \cite{cross1993pattern}. This is an interesting aspect which we leave for future investigations.

%{\color{red} Finally, generalized GL equations for many amplitudes could be derived for systems with many interacting species' populations (or different order parameters) $\phi_m(x,t)$, where the discrete index $m$ represents the trophic level to which they belong.}

\section*{Acknowledgments} 
S. G. acknowledges the support from Univeristy of Padova through the PhD fellowship within “Bando Dottorati di Ricerca”, funded by Fondazione Cassa di Risparmio di Padova e Rovigo. D. G. and A. M. acknowledge the support from University of Padova through “Excellence Project 2018” of Fondazione Cassa di Risparmio di Padova e Rovigo. We thank Samir Suweis for useful discussions. 

\section*{Author contributions}
Supervision and project conceptualization: A.M. and S.A. Theory and simulations: S.G. and D.G. All the authors discussed and analyzed results, and wrote the manuscript.

\appendix

%\begin{widetext}
%\newpage
%\pagestyle{empty}
%%%%%%%%%% Merge with supplemental materials %%%%%%%%%%
%\pagebreak

%\begin{center}\Large{Supplemental Material for ``Ginzburg-Landau amplitude equation for non-linear non-local models'' }\end{center}

\section{Derivation of Eq.~\eqref{eq-all-orders-with-space}}
\label{der-28}
In this section, we show the derivation to obtain the Eq.~\eqref{eq-all-orders-with-space}. We begin with substituting Eqs. \eqref{eq: time-sc}--\eqref{eq: nl-exp} into Eq. \eqref{eq: expand-eq} which gives
\begin{widetext}
\begin{align}
\epsilon^3 \dot{\varphi_1} + o\left(\epsilon^3\right)&=\epsilon\left(\mathcal{L}_{\textbf{p}_0} \varphi_1\right)+\epsilon^2\left[\mathcal{L}_{\textbf{p}_0} \varphi_2+C_{\textbf{q}_0}^{(2,0)}\varphi_1^2+C_{\textbf{q}_0}^{(1,1)}\varphi_1 (G_{\textbf{q}_0}*\varphi_1)+C_{\textbf{q}_0}^{(0,2)}\left(G_{\textbf{q}_0}*\varphi_1\right)^2\right]+ \notag\\
&+\epsilon^3 \bigg[\mathcal{L}_{\textbf{p}_0}\varphi_3+\delta \mathcal{L}_{\textbf{p}_0}\varphi_1+2C_{\textbf{q}_0}^{(2,0)}\varphi_1\varphi_2+C_{\textbf{q}_0}^{(1,1)}\left[\varphi_1(G_{\textbf{q}_0}*\varphi_2)+\varphi_2(G_{\textbf{q}_0}*\varphi_1)\right]+ \notag\\
&+2C_{\textbf{q}_0}^{(0,2)}(G_{\textbf{q}_0}*\varphi_1)(G_{\textbf{q}_0}*\varphi_2)+ C_{\textbf{q}_0}^{(3,0)}\varphi_1^3+C_{\textbf{q}_0}^{(2,1)}\varphi_1^2(G_{\textbf{q}_0}*\varphi_1)+ \notag\\
&+C_{\textbf{q}_0}^{(1,2)}\varphi_1(G_{\textbf{q}_0}*\varphi_1)^2+C_{\textbf{q}_0}^{(0,3)}(G_{\textbf{q}_0}*\varphi_1)^3 \bigg],
\label{eq: eq-all-orders}
\end{align}
\end{widetext}
where, for convenience, we have not written the $x,\xi,t$ dependence in $\varphi_i$.

Note that the expansion of Eq.~\eqref{eq: expand-eq} should also include all the contributions at different orders of $\epsilon$. Therefore, we have to also take into account the ones coming from the spatial scale separation. Using Eq.~\eqref{eq: space-der}, we can see that
\begin{equation}
\partial_x^2 \to \left(\partial_x +\epsilon\partial_\xi\right)^2=\partial_x^2 +2\epsilon\partial_x\partial_\xi+ \epsilon^2\partial_\xi^2,\label{pd-1}
\end{equation}
and this indicates how the Laplacian operator in the $\mathcal{L}_{\textbf{p}_0}$ given in Eq.~\eqref{eq: eq-all-orders}, transforms and operates on both $x$ and $\xi$ variables. 

Next ingredient we need in the following is the convolutions between the function $G_{\textbf{q}_0}(x)$ and $\varphi_i(x,\xi,\tau)$ that appear in Eq. \eqref{eq: eq-all-orders}: 
\begin{align}
(G_{\textbf{q}_0}*\varphi_i)(x,\xi,\tau)=\int_{-\infty}^{+\infty}~dy~G_{\textbf{q}_0}(x-y)\varphi_i(y,\xi',\tau)~dy,\label{eq: conv-1}
\end{align}
where $\xi=\epsilon x$ and $\xi'=\epsilon y$.  Following \cite{morgan2014swift}, we write the above integration \eqref{eq: conv-1}
 as \begin{align}
(G_{\textbf{q}_0}*\varphi_i)(x,\xi,\tau)= \int_{-\infty}^{+\infty}~dz~G_{\textbf{q}_0}(-z)~\varphi_i(x+z,\xi + \epsilon z,\tau).
\label{eq-150}
\end{align}
where we make a change in the integration variable from $x$ to $z=y-x$.

Expanding the above equation \eqref{eq-150} about the slow variable $\xi$, and integrating term by term yields
\begin{equation}
(G_{\textbf{q}_0}*\varphi_i)(x,\xi,\tau)=\sum_{n=0}^{\infty}\frac{\epsilon^n}{n!}(G_{\textbf{q}_0}*\varphi_i)_n,\label{eq: con-2}
\end{equation}
where, for brevity, we define \begin{equation} 
(G_{\textbf{q}_0}*\varphi_i)_n(x,\xi,\tau)=\int_{-\infty}^{+\infty}~dz~G_{\textbf{q}_0}(-z)z^n \frac{\partial ^n \varphi_i}{{\partial\xi}^n}(x+z,\xi,\tau).
\label{eq: expression-contribution-conv}
\end{equation}
With these considerations, the linear operator given in Eq.~\eqref{lin-op} can be rewritten as 
\begin{equation}
\mathcal{L}_{\textbf{p}_0}=\sum_{n=0}^{\infty}\epsilon^n\mathcal{L}_{\textbf{p}_0}^{(n)},
\end{equation} 
where \begin{align*}
\mathcal{L}_{\textbf{p}_0}^{(0)}\varphi_i(x,\xi,\tau)&=D_0\partial_x^2 \varphi_i(x,\xi,\tau)+C^{(1,0)}_{\textbf{q}_0}\varphi_i(x,\xi,\tau)+\notag \\
&+C^{(0,1)}_{\textbf{q}_0}\left(G_{\textbf{q}_0} * \varphi_i\right)_0(x,\xi,\tau), 
\end{align*}
\begin{align*}
\mathcal{L}_{\textbf{p}_0}^{(1)}\varphi_i(x,\xi,\tau)&=2D_0\partial_x \partial_\xi\varphi_i(x,\xi,\tau)+\notag \\ &+C^{(0,1)}_{\textbf{q}_0}(G_{\textbf{q}_0}*\varphi_i)_1(x,\xi,\tau),
\end{align*}
\begin{align*}
\mathcal{L}_{\textbf{p}_0}^{(2)}\varphi_i(x,\xi,\tau)&=D_0\partial_\xi^2\varphi_i(x,\xi,\tau)+\notag \\
&+\frac{1}{2}C^{(0,1)}_{\textbf{q}_0}(G_{\textbf{q}_0}*\varphi_i)_2(x,\xi,\tau),
\end{align*}
\begin{align*}
&\mathcal{L}_{\textbf{p}_0}^{(n\geq3)}\varphi_i(x,\xi,\tau)=\frac{1}{n!}C^{(0,1)}_{\textbf{q}_0}(G_{\textbf{q}_0}*\varphi_i)_n(x,\xi,\tau).
\end{align*}

Finally, we obtain Eq. \eqref{eq-all-orders-with-space} in which 
\begin{widetext}
\begin{align}
H_1(\textbf{p}_0,\varphi_1)&=\mathcal{L}_{\textbf{p}_0}^{(0)} \varphi_1,\nonumber\\
H_2(\textbf{p}_0,\varphi_1,\varphi_2)&=\mathcal{L}_{\textbf{p}_0}^{(0)} \varphi_2+C_{\textbf{q}_0}^{(2,0)}\varphi_1^2+C_{\textbf{q}_0}^{(1,1)}\varphi_1 (G_{\textbf{q}_0}*\varphi_1)_0+C_{\textbf{q}_0}^{(0,2)}\left(G_{\textbf{q}_0}*\varphi_1\right)_0^2+\mathcal{L}_{\textbf{p}_0}^{(1)} \varphi_1,\nonumber\\
H_3(\textbf{p}_0,\varphi_1,\varphi_2)&=\mathcal{L}_{\textbf{p}_0}^{(0)}\varphi_3+\delta \mathcal{L}_{\textbf{p}_0}^{(0)}\varphi_1+2C_{\textbf{q}_0}^{(2,0)}\varphi_1\varphi_2+C_{\textbf{q}_0}^{(1,1)}[\varphi_1(G_{\textbf{q}_0}*\varphi_2)_0+\varphi_2(G_{\textbf{q}_0}*\varphi_1)_0]+\notag \\
&+2C_{\textbf{q}_0}^{(0,2)}(G_{\textbf{q}_0}*\varphi_1)_0(G_{\textbf{q}_0}*\varphi_2)_0+ C_{\textbf{q}_0}^{(3,0)}\varphi_1^3+C_{\textbf{q}_0}^{(2,1)}\varphi_1^2(G_{\textbf{q}_0}*\varphi_1)_0C_{\textbf{q}_0}^{(1,2)}+\varphi_1(G_{\textbf{q}_0}*\varphi_1)_0^2+\notag \\
&+C_{\textbf{q}_0}^{(0,3)}(G_{\textbf{q}_0}*\varphi_1)_0^3+\mathcal{L}_{\textbf{p}_0}^{(2)} \varphi_1+\mathcal{L}_{\textbf{p}_0}^{(1)} \varphi_2+C_{\textbf{q}_0}^{(1,1)}\varphi_1 (G_{\textbf{q}_0}*\varphi_1)_1+2C_{\textbf{q}_0}^{(0,2)}\left(G_{\textbf{q}_0}*\varphi_1\right)_0 \left(G_{\textbf{q}_0}*\varphi_1\right)_1. \label{Seq: H-expressions}
\end{align}
\end{widetext}

\section{Derivation of Eq.~\eqref{sol-second order}}
\label{der-sec-order}
In this section, we present the detailed derivation to obtained the Eq.~\eqref{sol-second order}. To do so, we group the second order terms in Eq.~\eqref{eq-all-orders-with-space} by comparing the left and right-hand side, and we obtain

\begin{equation}
H_2(\textbf{p}_0,\varphi_1,\varphi_2)=0
\end{equation}
that can be rewritten extensively as
\begin{align}
\mathcal{L}_{\textbf{p}_0}^{(0)}\varphi_2&=-C_{\textbf{q}_0}^{(2,0)}\varphi_1^2-C_{\textbf{q}_0}^{(1,1)}\varphi_1 (G_{\textbf{q}_0}*\varphi_1)_0+\notag\\
&-C_{\textbf{q}_0}^{(0,2)}\left(G_{\textbf{q}_0}*\varphi_1\right)_0^2 -\mathcal{L}_{\textbf{p}_0}^{(1)} \varphi_1.
\label{Seq: second order}
\end{align}
In order to find the solution $\varphi_2$ we need to evaluate $(G_{\textbf{q}_0}*\varphi_1)_0$ and $\mathcal{L}_{\textbf{p}_0}^{(1)} \varphi_1$. Using Eqs. \eqref{eq: expression-contribution-conv} and \eqref{eq-phi-1}, we get
\begin{align}
(G_{\textbf{q}_0}*\varphi_1)_0(x,\xi,\tau)&=\int_{-\infty}^{+\infty}G_{\textbf{q}_0}(-z) \notag \\
&\times\bigg[A(\epsilon x,\tau)e^{ik_M(\textbf{p}_0) (x+z)}+\nonumber\\&+\bar{A}(\epsilon x, \tau)e^{-ik_M(\textbf{p}_0) (x+z)}\bigg] dz.
\end{align}

Thanks to the even nature of the function $G_{\textbf{q}_0}(z)$, we find
\begin{align}
(G_{\textbf{q}_0}*\varphi_1)_0(x,\xi,\tau)&=\tilde{G}_{\textbf{q}_0}(k_M(\textbf{p}_0))\bigg[A(\xi,\tau)e^{ik_M(\textbf{p}_0) x}+\notag \\
&+\bar{A}(\xi, \tau)e^{-ik_M(\textbf{p}_0) x}\bigg]\nonumber\\&=\tilde{G}_{\textbf{q}_0}(k_M(\textbf{p}_0)) \varphi_1(x,\xi,\tau).\label{op-1}
\end{align}

Let us now evaluate $\mathcal{L}_{\textbf{p}_0}^{(1)} \varphi_1$. Doing some algebra, we get
\begin{equation}
\mathcal{L}_{\textbf{p}_0}^{(1)} \varphi_1=C^{(0,1)}_{\textbf{q}_0}(G_{\textbf{q}_0}*\varphi_1)_1+2D_0\partial_x \partial_\xi\varphi_1,
\end{equation}
where
\begin{align}
(G_{\textbf{q}_0}*\varphi_1)_1(x,\xi,\tau)&= (\partial_\xi A)(\epsilon x,\tau)e^{ik_M(\textbf{p}_0) (x)}I+\notag \\
&+(\partial_\xi \bar{A})(\epsilon x,\tau)e^{-ik_M(\textbf{p}_0) (x)}\bar{I},
\label{Seq: conv-G-phi_1-order-1}
\end{align}
in which the integral 
\begin{align}
I=\int_{-\infty}^{+\infty}G_{\textbf{q}_0}(-z)ze^{ik_M(\textbf{p}_0) z}dz=-i \tilde{G}'(k_M(\textbf{p}_0)),
\label{Seq: conv-G-phi_1-order-1-int-I}
\end{align}
and $\bar I$ is its complex conjugate. Therefore, $\mathcal{L}_{\textbf{p}_0}^{(1)} \varphi_1$ becomes
\begin{align}
\mathcal{L}_{\textbf{p}_0}^{(1)} \varphi_1&=i\partial_\xi A(\xi,\tau)e^{ik_M(\textbf{p}_0) x}\underbrace{\lambda_{\textbf{p}_0}'(k_M(\textbf{p}_0))}_{=0}+\notag\\
&+i\partial_\xi \bar{A}(\xi,\tau)e^{-ik_M(\textbf{p}_0) x}\underbrace{\lambda_{\textbf{p}_0}'(-k_M(\textbf{p}_0))}_{=0}=0.
\label{op-3}
\end{align}

Using Eqs.~ \eqref{op-1}, \eqref{Seq: conv-G-phi_1-order-1}, and \eqref{op-3} in Eq.~\eqref{Seq: second order}, we finally get 
\begin{equation}
\mathcal{L}_{\textbf{p}_0}^{(0)} \varphi_2=\Sigma_{\textbf{p}_0} \varphi_1^2,
\label{Seq: second order-reduced}
\end{equation}
where we define the coefficient $\Sigma_{\textbf{p}_0}$ as
\begin{align}
\Sigma_{\textbf{p}_0}&=-C_{\textbf{q}_0}^{(2,0)}-C_{\textbf{q}_0}^{(1,1)}\tilde{G}_{\textbf{q}_0}(k_M(\textbf{p}_0))+\notag \\
&-C_{\textbf{q}_0}^{(0,2)}\tilde{G}_{\textbf{q}_0}(k_M(\textbf{p}_0))^2.
\end{align}

Clearly, Eq.~\eqref{Seq: second order-reduced} satisfies the Fredholm's alternative since $\varphi_1^2 \not \in  ker\left(\mathcal{L}_{\textbf{p}_0}^{(0)}\right)$. In fact, the right-hand side of Eq.~\eqref{Seq: second order-reduced}  is orthogonal to $\varphi_1$, and therefore, using Fredholm's alternative, Eq. \eqref{Seq: second order} admits a bounded solution. Thus, using Eq. \eqref{eq-phi-1} in \eqref{Seq: second order-reduced}, we obtain the solution $\varphi_2(x,\xi,\tau)$  and it is shown in Eq.~\eqref{sol-second order}.

\section{Derivation of Eq.~\eqref{eq: amplitude-eq}: the GL amplitude equation}
\label{third-order}
Here, we obtain the GL amplitude equation shown in Eq.~\eqref{eq: amplitude-eq}. In the following, we compare the terms of third order in $\epsilon$ in the two sides of  expansion \eqref{eq-all-orders-with-space}. Therefore, we get

\begin{equation}
\dot{\varphi}_1=H_3(\textbf{p}_0,\varphi_1,\varphi_2)
\end{equation}
that can be recast as
\begin{widetext}
\begin{align}
-\mathcal{L}_{\textbf{p}_0}^{(0)}\varphi_3&=-\dot{\varphi}_1+\delta \mathcal{L}_{\textbf{p}_0}^{(0)}\varphi_1+2C_{\textbf{q}_0}^{(2,0)}\varphi_1\varphi_2+C_{\textbf{q}_0}^{(1,1)}\left[\varphi_1(G_{\textbf{q}_0}*\varphi_2)_0+\varphi_2(G_{\textbf{q}_0}*\varphi_1)_0\right]+\notag \\
&+2C_{\textbf{q}_0}^{(0,2)}(G_{\textbf{q}_0}*\varphi_1)_0(G_{\textbf{q}_0}*\varphi_2)_0+C_{\textbf{q}_0}^{(3,0)}\varphi_1^3+C_{\textbf{q}_0}^{(2,1)}\varphi_1^2(G_{\textbf{q}_0}*\varphi_1)_0+C_{\textbf{q}_0}^{(1,2)}\varphi_1(G_{\textbf{q}_0}*\varphi_1)_0^2+\notag \\
&+C_{\textbf{q}_0}^{(0,3)}(G_{\textbf{q}_0}*\varphi_1)_0^3 +C_{\textbf{q}_0}^{(1,1)}\varphi_1(G_{\textbf{q}_0}*\varphi_1)_1+2C_{\textbf{q}_0}^{(0,2)}\left(G_{\textbf{q}_0}*\varphi_1\right)_0 \left(G_{\textbf{q}_0}*\varphi_1\right)_1+\mathcal{L}_{\textbf{p}_0}^{(2)} \varphi_1+\mathcal{L}_{\textbf{p}_0}^{(1)} \varphi_2.
\label{Seq: third order}
\end{align}
\end{widetext}

We substitute the expression of $(G_{\textbf{q}_0}*\varphi_2)_0$ [following Eqs.~\eqref{eq: expression-contribution-conv} and \eqref{sol-second order}], $\delta \mathcal{L}_{\textbf{p}_0}^{(0)}\varphi_1$, and $\mathcal{L}_{\textbf{p}_0}^{(2)} \varphi_1$:
\begin{widetext}
\begin{align}
(G_{\textbf{q}_0}*\varphi_2)_0(x,\xi,\tau)&=\Sigma_{\textbf{p}_0}\bigg\lbrace\frac{\tilde{G}_{\textbf{q}_0}(2k_M(\textbf{p}_0))}{\lambda_{\textbf{p}_0}(2k_M(\textbf{p}_0))}\big[ A^2(\xi,\tau) e^{2ik_M(\textbf{p}_0) x}
+\bar{A}^2(\xi,\tau)e^{-2ik_M(\textbf{p}_0) x}\big]+\frac{2 \tilde{G}_{\textbf{q}_0}(0)}{\lambda_{\textbf{p}_0}(0)}\vert A(\xi,\tau) \vert ^2\bigg\rbrace +\notag \\
&+\tilde{G}_{\textbf{q}_0}(k_M(\textbf{p}_0))\Lambda(x,\xi,\tau),\label{eqn-53} 
\end{align}
\end{widetext}
\begin{align}
\delta \mathcal{L}_{\textbf{p}_0}^{(0)}\varphi_1&=\hat{v}\cdot\left( \vec{\nabla}_{\textbf{p}}\mathcal{L}_{\textbf{p}}^{(0)}\right)\vert_{\textbf{p}=\textbf{p}_0}\varphi_1\equiv\bar{\lambda}_M \varphi_1,\label{eqn-54}
\end{align}
\begin{align}
\mathcal{L}_{\textbf{p}_0}^{(2)} \varphi_1&=\frac{1}{2}C^{(0,1)}_{\textbf{q}_0}(G_{\textbf{q}_0}*\varphi_1)_2+D_0\partial_\xi^2\varphi_1, \label{eqn-55}
\end{align}
in Eq. \eqref{Seq: third order}, where
\begin{align}
(G_{\textbf{q}_0}*\varphi_1)_2(x,\xi,\tau)&=-\tilde{G}''(k_M(\textbf{p}_0))\partial_\xi^2 A(\xi,\tau))e^{ik_M(\textbf{p}_0)x}+\notag \\
&-\tilde{G}''(-k_M(\textbf{p}_0))\partial_\xi^2 \bar{A}(\xi,\tau))e^{-ik_M(\textbf{p}_0)x}.
\end{align}
Notice that in arriving the above form of $(G_{\textbf{q}}*\varphi_1)_2(x,\xi,\tau)$  we have used  the same strategy as in Eq. \eqref{Seq: conv-G-phi_1-order-1}. Thus, Eq. \eqref{eqn-55} becomes
\begin{align}
\mathcal{L}_{\textbf{p}_0}^{(2)} \varphi_1&=-\frac{1}{2}\lambda_{\textbf{p}_0}''(k_M(\textbf{p}_0))\partial_\xi^2 A(\xi,\tau))e^{ik_M(\textbf{p}_0)x}+\notag \\
&-\frac{1}{2}\lambda_{\textbf{p}_0}''(-k_M(\textbf{p}_0))\partial_\xi^2 \bar{A}(\xi,\tau))e^{-ik_M(\textbf{p}_0)x}. \label{lp0-1}
\end{align}

Finally, we substitute Eqs. \eqref{eqn-53}, \eqref{eqn-54}, \eqref{lp0-1}, and $\varphi_1$ from Eq. \eqref{eq-phi-1} in Eq.~\eqref{Seq: third order}. Since $\varphi_3$ has to be bounded, the right-hand side of Eq.~\eqref{Seq: third order} must be orthogonal to $\varphi_1$ (Fredholm's alternative). Therefore, setting the coefficients of $e^{ik_M(\textbf{p}_0)x}$ in Eq. \eqref{Seq: third order} equal to zero while noticing that $
\mathcal{L}_{\textbf{p}_0}^{(1)} \varphi_2+C_{\textbf{q}_0}^{(1,1)}\varphi_1 (G_{\textbf{q}_0}*\varphi_1)_1+2C_{\textbf{q}_0}^{(0,2)}\left(G_{\textbf{q}_0}*\varphi_1\right)_0 \left(G_{\textbf{q}_0}*\varphi_1\right)_1$
does not have any term proportional to $e^{ik_M(\textbf{p}_0)x}$, we obtain the GL amplitude equation as shown in Eq.~\eqref{eq: amplitude-eq},
where $\bar{\lambda}_M$ is given in \eqref{Seq: exp-lambda-max-tot} and the coefficient $\alpha$ has the following form:
\begin{widetext}
\begin{align}
\alpha&=-\bigg\lbrace2\Sigma_{\textbf{p}_0} C_{\textbf{q}_0}^{(2,0)}\bigg[\frac{2}{\lambda_{\textbf{p}_0}(0)}+\frac{1}{\lambda_{\textbf{p}_0}(2k_M(\textbf{p}_0))}\bigg]+\Sigma_{\textbf{p}_0} C_{\textbf{q}_0}^{(1,1)}\bigg[2\frac{\tilde{G}_{\textbf{q}_0}(0)+\tilde{G}_{\textbf{q}_0}(k_M(\textbf{p}_0))}{\lambda_{\textbf{p}_0}(0)}+ \notag \\
&+\frac{\tilde{G}_{\textbf{q}_0}(k_M(\textbf{p}_0))+\tilde{G}_{\textbf{q}_0}(2k_M(\textbf{p}_0))}{\lambda_{\textbf{p}_0}(2k_M(\textbf{p}_0))}\bigg]+2\Sigma_{\textbf{p}_0} C_{\textbf{q}_0}^{(0,2)}\tilde{G}_{\textbf{q}_0}(k_M(\textbf{p}_0))\left[\frac{2\tilde{G}_{\textbf{q}_0}(0)}{\lambda_{\textbf{p}_0}(0)}+\frac{\tilde{G}_{\textbf{q}_0}(2k_M(\textbf{p}_0))}{\lambda_{\textbf{p}_0}(2k_M(\textbf{p}_0))}\right]+ \notag \\
&+3 C_{\textbf{q}_0}^{(3,0)}+3C_{\textbf{q}_0}^{(2,1)}\tilde{G}_{\textbf{q}_0}(k_M(\textbf{p}_0))+3C_{\textbf{q}_0}^{(1,2)}(\tilde{G}_{\textbf{q}_0}(k_M(\textbf{p}_0)))^2+3C_{\textbf{q}_0}^{(0,3)}(\tilde{G}_{\textbf{q}_0}(k_M(\textbf{p}_0)))^3\bigg\rbrace.
\label{Seq: alpha-def}
\end{align}
\end{widetext}

\section{Particular solutions of the GL amplitude equation}\label{P-S}
In this section, we present two interesting analytical solutions of the GL amplitude equation \eqref{eq: amplitude-eq}. 

Let us substitute the complex amplitude $A(\xi,\tau)$:  
\begin{equation}
A(\xi,\tau)=\vert A(\xi,\tau)\vert e^{i\theta(\xi,\tau)}
\label{Seq: amp-exp-expression}
\end{equation}
where both $\vert A(\xi,\tau)\vert$ and $\theta(\xi,\tau)$ are real functions of $\xi$ and $\tau$, in Eq.~\eqref{eq: amplitude-eq}. Separating the real and imaginary parts, we obtain a set of coupled differential equations for the modulus $|A(\xi,\tau)|$ and the phase of the amplitude $\theta(\xi,\tau)$: 
\begin{align}
\partial_\tau \vert A \vert&=\bar{\lambda}_M \vert A \vert-\alpha \vert A \vert ^3 +\notag \\ &+\frac{1}{2}\big|\lambda_{\textbf{p}_0}''(k_M(\textbf{p}_0))\big|\left[ \partial^2_\xi \vert A \vert - \vert A \vert (\partial_\xi \theta)^2\right],\label{Seq: real-im-amp-eq-1} \\
\vert A \vert\partial_\tau \theta &= \frac{1}{2}\big|\lambda_{\textbf{p}_0}''(k_M(\textbf{p}_0))\big|\left[2(\partial_\xi \vert A \vert)(\partial_\xi \theta)+\vert A \vert\partial^2_\xi \theta\right],
\label{Seq: real-im-amp-eq-2}
\end{align}
where, for convenience, we have dropped the arguments in both $|A(\xi,\tau)|$ and $\theta(\xi,\tau)$.

It is difficult to obtain the solution of above coupled differential for a generic initial condition. Nonetheless, for some particular initial conditions, the exact solution can be obtained. As a first example, we consider an initial homogeneous condition, i.e.,
\begin{equation}
A(\xi,0)\equiv A_0 e^{i \theta_0}, \label{Seq: homog-IC}
\end{equation}
where both $A_0$ and $\theta_0$ are independent of $\xi$. Therefore, the solution in this case can be obtained as 
\begin{align}
\vert A(\xi,\tau) \vert&=\frac{A_0 \sqrt{\bar{\lambda}_M} \exp{\left(\bar{\lambda}_M \tau\right)}}{\sqrt{\bar{\lambda}_M+A_0^2 \alpha \left[ \exp{\left(2\bar{\lambda}_M \tau\right)}-1 \right]}}, \label{sol-hom-1} \\
\theta(\xi,\tau)&=\theta_0, \label{sol-hom-2}  
\end{align}
 and they satisfy both Eq.~\eqref{Seq: real-im-amp-eq-1} and \eqref{Seq: real-im-amp-eq-2} and the initial condition Eq.~\eqref{Seq: homog-IC}. Thus, for a given initial homogeneous condition, the GL amplitude equation predicts the amplitude to be homogeneous where only the modulus $|A|$ evolves with time $\tau$.

To obtain a spatial solution of the amplitude equation, we again consider an initial homogeneous condition for the phase, i.e., $\theta(\xi,0)\equiv \theta_0$. Thus, the equation for the modulus of the amplitude reduces to 
\begin{equation}
\partial_\tau \vert A \vert=\bar{\lambda}_M \vert A \vert-\alpha \vert A \vert ^3  +\frac{1}{2}\vert\lambda_{\textbf{p}_0}''(k_M(\textbf{p}_0)) \vert\partial^2_\xi \vert A \vert 
\label{Seq: amp-tanh-1}
\end{equation}
A steady solution $\vert A_{st}(\xi)\vert$ of above Eq.~\eqref{Seq: amp-tanh-1} can be obtained by setting the left hand side of Eq.~\eqref{Seq: amp-tanh-1} to $0$, and we get
\begin{equation}
\vert A_{st}(\xi) \vert=\pm\sqrt{\frac{\bar{\lambda}_M}{\alpha}}\tanh \left[\xi\sqrt{\frac{\bar{\lambda}_M}{\big|\lambda''(k_M(\textbf{p}_0))\big|}}\right].
\label{Seq: amp-tanh-2}
\end{equation}
as one possible solution, as shown in Ref.~\cite{Hoyle}.

Since $\vert A_{st}(\xi) \vert$ must be non-negative, a solution that satisfies this condition can be constructed as 
\begin{equation}\label{sol:tanh}
\vert A_{st}(\xi) \vert = \sqrt{\frac{\bar{\lambda}_M}{\alpha}}\tanh \left[\vert \xi \vert \sqrt{\frac{\bar{\lambda}_M}{\big|\lambda''(k_M(\textbf{p}_0))\big|}}\right]. 
% \begin{cases}
% -\sqrt{\frac{\bar{\lambda}_M}{\alpha}}\tanh \left[\xi\sqrt{-\frac{\bar{\lambda}_M}{\lambda''(k_M(\textbf{p}_0))}}\right] \qquad \text{if } \xi<0\\
% 0\hspace{5.27cm} \text{if } \xi=0\\
% \sqrt{\frac{\bar{\lambda}_M}{\alpha}}\tanh \left[\xi\sqrt{-\frac{\bar{\lambda}_M}{\lambda''(k_M(\textbf{p}_0))}}\right] \hspace{1cm} \text{if } \xi>0
% \end{cases}
\end{equation}
In the above solution, we consider both solutions \eqref{Seq: amp-tanh-2} depending on the sign of the variable $\xi$ and introduce a defect at $\xi=0$, where the amplitude becomes zero. In fact, this solution also satisfies the amplitude equation everywhere except at the defect where it changes the behavior passing from one to the other solution displayed in Eq.~\eqref{Seq: amp-tanh-2}. 

It is possible to show analytically that the homogeneous solution of Eq.~\eqref{eq: amplitude-eq} is linearly stable while the steady spatial one \eqref{Seq: amp-tanh-2} is locally linearly unstable. %{\color{red}In other words, the latter is metastable which can be tested numerically up to  a finite observation time. Nonetheless, to display the steady behaviour also at longer observation time, the perfect tuning of the model parameters in the numerics would be required, which is difficult to handle in the model considered here.}
In other words, the numerical spatial solution is a good approximation of the analytical prediction only up to a finite observation time. Indeed, because of numerical inaccuracies, at larger time scales the profile will inevitably fall into the basin of attraction of the stationary stable solution.
%{\color{cyan}In other words, the latter is metastable which can be tested numerically up to  a finite observation time. Nonetheless, the perfect tuning of the parameters in the numerics can display the steady behaviour at longer observation time, which is difficult to handle in the model considered here.} 
%{\color{red}Nonetheless, to display the steady behaviour also at longer observation time, the perfect tuning of the model parameters in the numerics would be required, but this is a difficult task to handle  given the model considered here.}
%{\color{red}Hence, the latter is numerically metastable, i.e. Eq.~\eqref{Seq: amp-tanh-2} can be observed only up to a finite time-scale when we try to check its validity performing numerical test because of the impossibility of a perfect tuning of the parameters used in the numerical analysis.}

%In the following section, we compare the theoretical prediction of the amplitude with the numerical simulation of discrete \textcolor{red}{NLF equation} model.

\section{Numerical Methods}\label{num-met}
In this section, we discuss the method of numerical simulation to verify the analytical prediction of the amplitude equation (32) of the main text.
As an example, we consider the discrete nonlocal Fisher equation. To do so we consider a one dimensional line where the spatial variable $x$ ranges from $-L$ to $L$. Then we discretize the space creating a lattice introducing the discrete spatial variable $x_i$ defined as follows 
\begin{align}
x_i=-L +i~dx \quad \text{where} \quad i=1, \dots, N,
\end{align}
with $x_N=x_0$ [i.e., periodic boundary condition (PBC)]. In the above equation, $dx=2L/N$ is the uniform spacing. 
\begin{figure*}
    \centering
    \includegraphics[width=.45\textwidth]{amplitude-vs-tau-letter-panels-leg-first-order.pdf}~~~~
    \includegraphics[width=.45\textwidth]{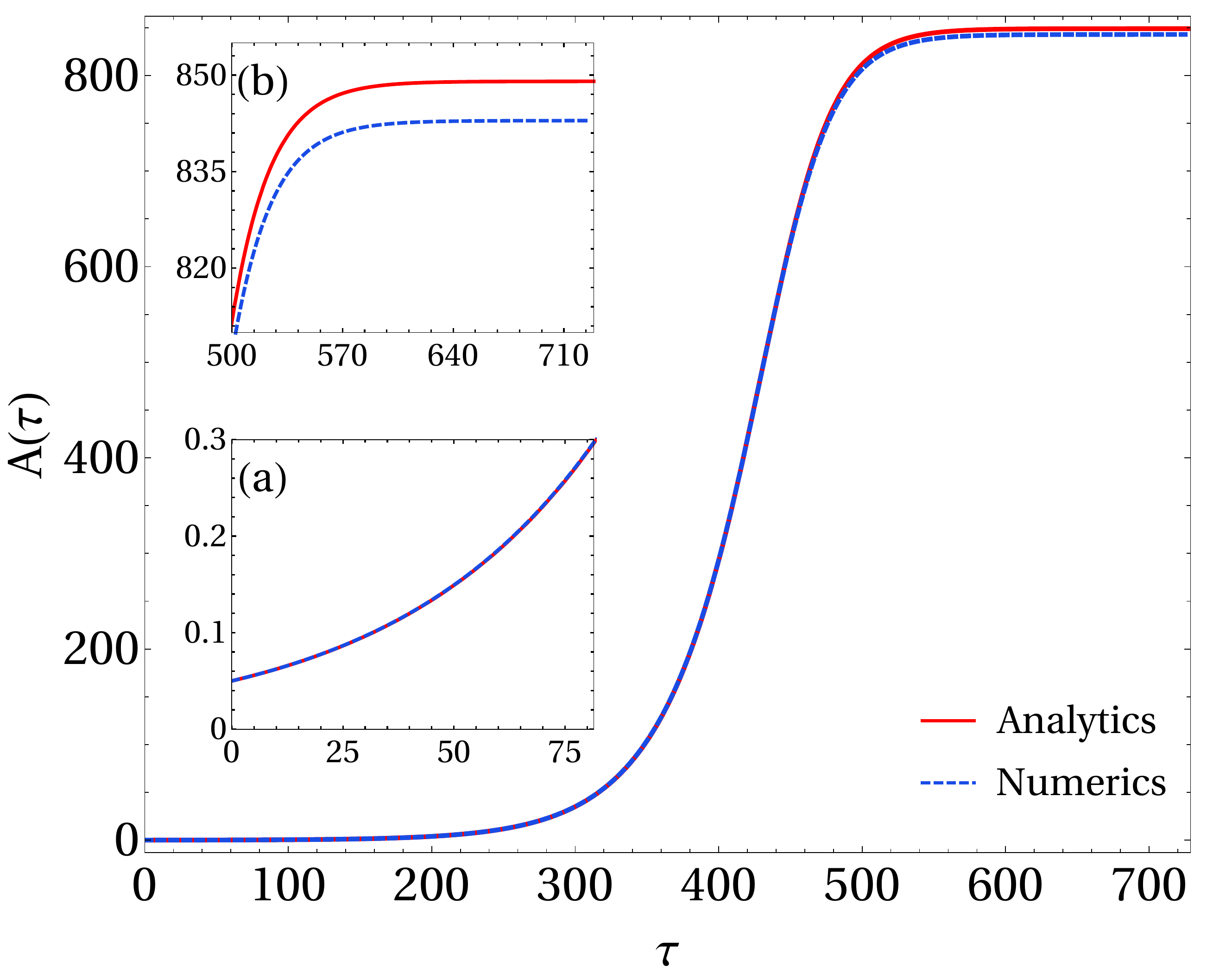}
    \caption{Comparison between theoretical prediction of the GL amplitude equation \eqref{eq: amplitude-eq} with the initial condition $A(\xi_i,0)=\vert A(\xi_i,0)\vert=A_0=0.05$  (solid red line) and the amplitude obtained from the numerical simulation (blue dashed line) for the discrete nonlocal F-KPP equation where the system is initialized in the state $\phi_i(\xi_i, 0)=\phi(x_i,\xi_i,0)=\phi_{\textbf{q}}^{(0)}+2\epsilon A_0 \cos\left[k_M(\textbf{p}_0) x_i \right]$. In the numerical implementation, we take $N=3060$ species equispaced along a ring of length $2L=3$, and these are interacting among each others with an interaction kernel given by $G_\textbf{q}(z)=\exp{\left(-\frac{\vert z \vert}{R}\right)}-b~\exp{\left(-\frac{\vert z \vert}{\beta R}\right)}$. In the left panel, the amplitude is extracted from the numerical simulation exploiting Eq.~\eqref{dis-amp} whereas in the right panel, we employ the truncated series \eqref{eq: perturb} up to second order to estimate the amplitude from the same numerical simulation. Insets in the two plots show the zooming of the curves up to a particular range of time $\tau$. Both plots are shown for fixed sets of parameters $\textbf{p}$ and $\textbf{p}_0$. In particular, here we set $R=0.1$, $\beta=0.5851$, $b=0.6$, $a=10^{-4}$, and $D=10^{-8}$. To compute the coefficients of Eq.~\eqref{eq: amplitude-eq} we used the set $\textbf{p}_0$ in which we tuned $\beta$ leaving the other parameters fixed.}
    \label{fig: hom-amp}
\end{figure*}

\begin{figure*}
    \centering
    \includegraphics[width=.45\textwidth]{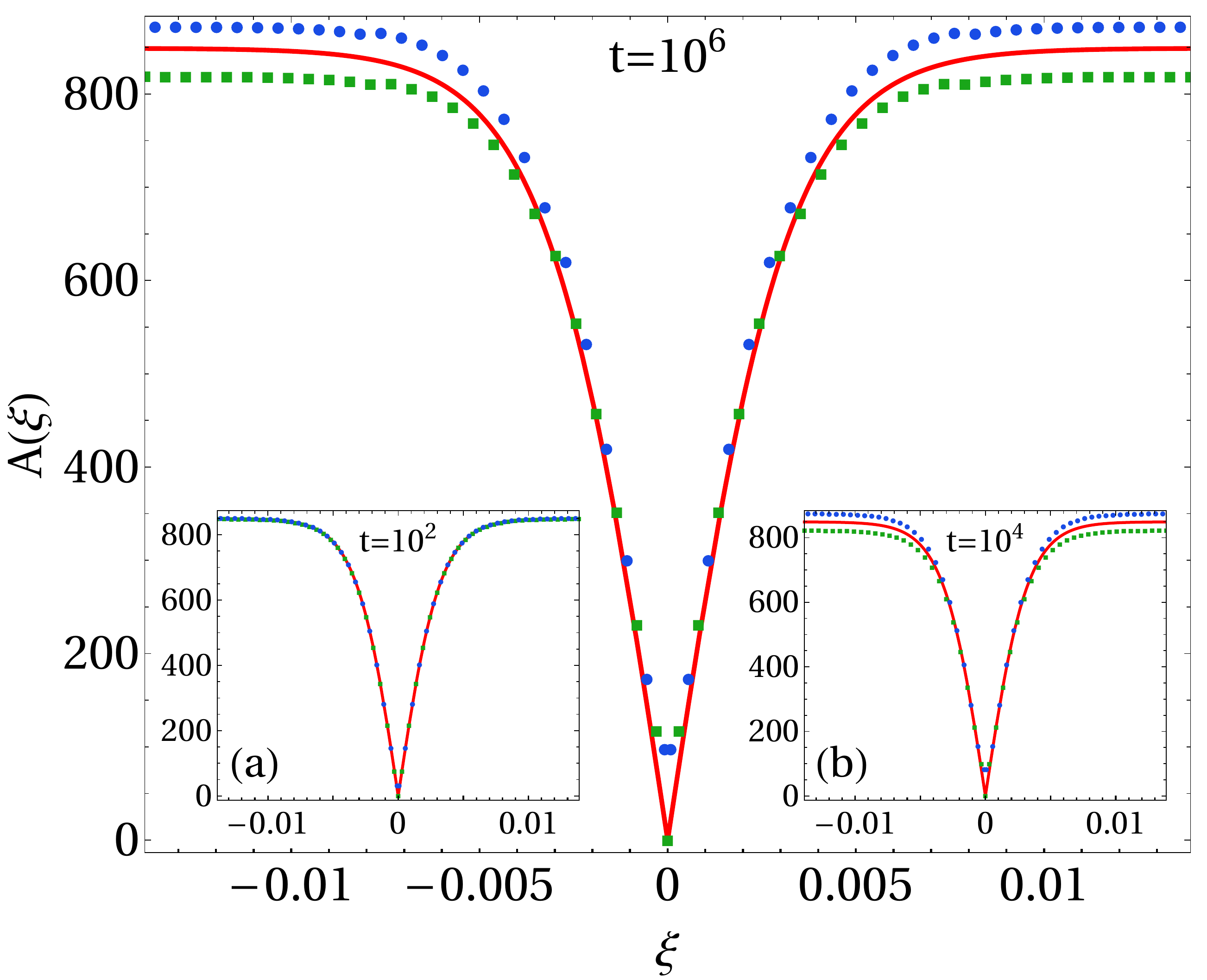}~~~~
    %{amp-spatial-dep-first-order.pdf}~~~~
    \includegraphics[width=.45\textwidth]{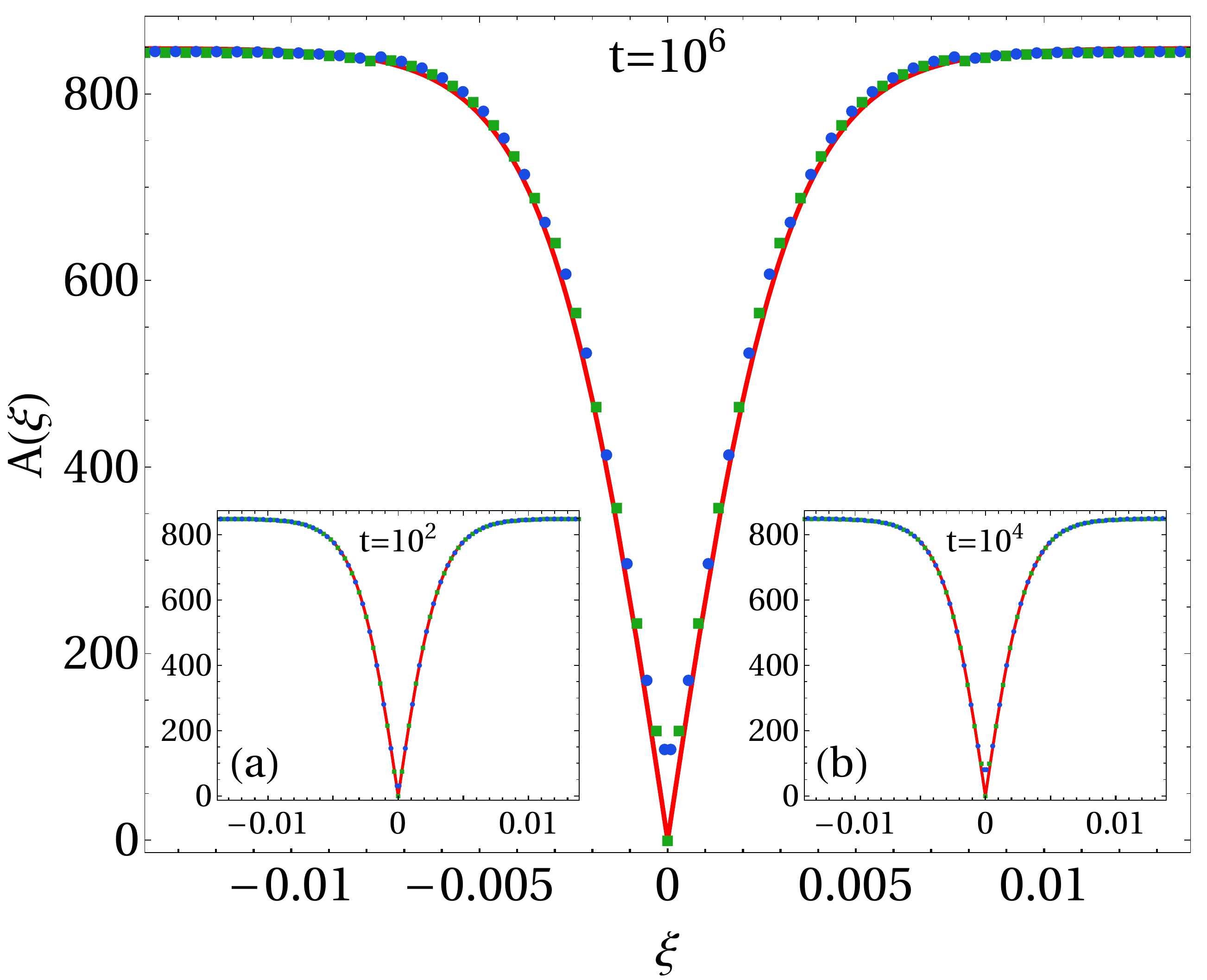}%{amp-spatial-dep-sec-order.pdf}
    \caption{Comparison of theoretical prediction of the GL amplitude equation \eqref{eq: amplitude-eq} (solid red line) with the numerical simulation (blue circles and green squares) for the discrete nonlocal F-KPP equation using the defective steady solution $A_{st}(\xi)$ \eqref{sol:tanh} and $\phi_i(\xi_i, 0)=\phi(x_i,\xi_i,0)=\phi_{\textbf{q}}^{(0)}+2\epsilon A_{st}(\xi_i) \cos\left[k_M\textbf{p}_0) x_i \right]$, respectively, as initial conditions. From the numerical integration of the discrete CLV dynamics, we extract the envelope of the pattern  using its local maxima (circles) and minima (squares). In the left panel, the amplitude is extracted from the numerical simulation exploiting Eq.~\eqref{dis-amp} whereas in the right panel, we use the truncated series \eqref{eq: perturb} up to second order to estimate the amplitude from the same numerical simulation. We show in the main plots the comparison at $t=10^6$ of the discrete nonlocal F-KPP equation, while in the insets the comparison is displayed at $t=10^2$ (a) and $t=10^4$ (b). Clearly, we can see that when we consider the higher-order contribution the agreement improves at larger time. The simulated dynamics, including the interaction kernel and the sets of parameter $\textbf{p}$ and $\textbf{p}_0$ used, is the same one presented in the caption of Figure~\ref{fig: hom-amp}, where the initial condition has been changed.
    }
    \label{fig: tanh-amp}
\end{figure*}

The dynamics described by the discrete nonlocal Fisher-KPP equation reads as
\begin{widetext}
\begin{equation}
\partial_t \phi_i(t)=\phi_i(t)\bigg[1-a\sum_{j=1}^N G_\textbf{q}\left(\min\lbrace \vert i-j \vert dx, 2L-\vert i-j \vert dx \rbrace\right)\phi_j(t)\bigg]+D \Delta \phi_i(t).
\label{Seq: LV-discrete}
\end{equation}
\end{widetext}
where the kernel respects PBC. The above equations \eqref{Seq: LV-discrete} are supplemented with  initial conditions $\phi_i(t=0)$ which we will discuss later. 
 
In the above Eq.~\eqref{Seq: LV-discrete}, the subscript  $i$  corresponds to  $i$-th position along the lattice, $\phi_i(t)$ is the value of the field at that position at time $t$ and the discrete Laplacian operator $\Delta$  acting on the field $\phi_i$ is defined as
\begin{align*}
\Delta \phi_i=\frac{\phi_{i-1}-2\phi_i+\phi_{i+1}}{dx^2}.
\end{align*}

The homogeneous and stationary solution corresponding to Eq. \eqref{Seq: LV-discrete} is given by
\begin{align}
\phi_\textbf{q}^{(0)}&=\frac{1}{a \sum_{j=1}^N G_\textbf{q}\left(\min\lbrace \vert i-j \vert~ dx, 2L-\vert i-j \vert~dx \rbrace\right)}\nonumber\\
&=\frac{1}{2 a\sum _{j=1}^{\frac{N}{2}-1} G_\textbf{q}(j~dx)+a~G_\textbf{q}(L)+a~G_\textbf{q}(0)}.
\end{align}

Now, to understand the stability of $\phi_\textbf{q}^{(0)}$, we substitute $\phi_j(t)\equiv \phi_\textbf{p}^{(0)}+\delta e^{\lambda_\textbf{p}(k_n)t+ik_n x_j}+c.c.$, where $0<\delta \ll 1$ and $k_n=n\frac{\pi}{L}$ with $n$ being an integer,  in Eq. \eqref{Seq: LV-discrete}. Therefore, we obtain the following dispersion relation (up to a linear order in $\delta$)
\begin{align}
\lambda_\textbf{p}(k_n)=-\frac{\tilde{g}_\textbf{q}(k_n)}{\tilde{g}_\textbf{q}(0)}+2D\frac{\cos \left(k_n~dx \right)-1}{dx^2}, \label{dis-lm}
\end{align}
where we have introduced the discrete Fourier transform as
\begin{align}
\tilde{g}_\textbf{q}(k_n)&=2 \sum _{j=1}^{\frac{N}{2}-1} \cos \left(k_n~j~dx\right) G_\textbf{q}(j~dx )+\notag \\
&+(-1)^n G_\textbf{q}(L)+G_\textbf{q}(0).
\label{Seq: trasf-disc-four}
\end{align}

In the following, we describe the recipe to obtain the amplitude of the pattern formed near the critical hypersurface $\mathcal{M}$ (Fig.~\ref{fig:phase-diag-and-lambda}) by numerical simulating Eq. \eqref{Seq: LV-discrete}. We stress that the theoretical prediction of amplitude equation [see Eq.~\eqref{eq: amplitude-eq}] does not get affected for the above discussed model. In this case, we just replace the Fourier transform with its discrete counterpart \eqref{Seq: trasf-disc-four}. 

%\sout{First, we consider a point $\textbf{p}$ in the pattern forming region (See Fig. 1 of the main text) and find the value of $k_M(\textbf{p})$ using Eq. \eqref{dis-lm}, where $k_M(\textbf{p})=%\underset{k_n}{\max}\{\lambda_\textbf{p}(k_n)\} $\max_{k_n}\{\lambda_\textbf{p}(k_n)\}$ and the point $\textbf{p}_0$, that lies on $\mathcal{M}$ around which we perform perturbation, is chosen as discussed in the main text.} 
First, we consider a point $\textbf{p}$ in the pattern forming region (See Fig. \ref{fig:phase-diag-and-lambda}) and find the value of $\lambda_M$ using Eq. \eqref{dis-lm}, where $\lambda_M=%\underset{k_n}{\max}\{\lambda_\textbf{p}(k_n)\} $
\max_{k_n}\{\lambda_\textbf{p}(k_n)\}$. Then we take the point $\textbf{p}_0$, that lies on $\mathcal{M}$ around which we perform the expansion as discussed in the main text, and we compute $k_M(\textbf{p}_0)$ and the coefficients appearing in Eq.~\eqref{eq: amplitude-eq} of the main text.

We note that in general for the continuous model shown in Eq.~\eqref{eq: general-eq} of the main text, the analytical solution of the dynamics [using solution of Eq.~\eqref{eq: amplitude-eq} given initial conditions,  and  Eq.~\eqref{eq: perturb}] can be written as (up to first order in $\epsilon$)
\begin{align}
\phi(x,\xi,\tau)&\approx\phi_{\textbf{q}}^{(0)}+\epsilon\varphi_1(x,\xi,\tau)\nonumber\\
&\approx\phi_{\textbf{q}}^{(0)}+2\epsilon \vert A(\xi, \tau) \vert \cos \left[k_M(\textbf{p}_0) x + \theta(\xi,\tau)\right],
\end{align}  
where $A(\xi,\tau)=|A(\xi,\tau)| e^{i\theta(\xi,\tau)}$. Therefore, the analogous discrete version of the above solution is 
\begin{align}\label{dis-amp}
\phi_i(\xi_i, \tau)&=\phi(x_i,\xi_i,\tau) \notag \\
&\approx\phi_{\textbf{q}}^{(0)}+2\epsilon \vert A(\xi_i, \tau) \vert \cos \left[k_M(\textbf{p}_0) x_i + \theta(\xi_i,\tau)\right],
\end{align}
where $x_i$ corresponds to discrete spatial location of the $i$-th species.

Here we aim to compare the amplitude given in the Eq.~\eqref{dis-amp} with the numerical simulation. To do so, we use the same initial and boundary conditions imposed on the solution~\eqref{dis-amp}. Finally, we verify the analytical prediction for growth of the amplitude for two different initial conditions given in Eqs.~\eqref{sol-hom-1}, \eqref{sol-hom-2}, and \eqref{sol:tanh} in Figs. \ref{fig: hom-amp} and \ref{fig: tanh-amp}.

%\end{widetext}

\end{document}